\documentstyle[twocolumn,aps]{revtex}
\def\Journal#1#2#3#4{{#1} {\bf #2}, #3 (#4)}
\input{epsfig.sty}

\def\NPA{{\rm Nucl. Phys.} A}
\def\NPB{{\rm Nucl. Phys.} B}
\def\PLB{{\rm Phys. Lett.}  B}
\def\PRL{\rm Phys. Rev. Lett.}
\def\PRD{{\rm Phys. Rev.} D}

\def\la{\langle}
\def\ra{\rangle}
\def\al{\alpha}
\def\be{\begin{equation}}
\def\ee{\end{equation}}
\def\bea{\begin{eqnarray}}
\def\eea{\end{eqnarray}}
\def\lsim{\mathrel{\rlap{\lower4pt\hbox{\hskip1pt$\sim$}}
    \raise1pt\hbox{$<$}}}         
\def\gsim{\mathrel{\rlap{\lower4pt\hbox{\hskip1pt$\sim$}}
    \raise1pt\hbox{$>$}}}

\begin{document}
\title{ Light-front quark model analysis of rare
$B\to K\ell^+\ell^-$ decays}
\author{ Ho-Meoyng Choi$^{a}$, Chueng-Ryong Ji$^{b}$
and L.S. Kisslinger$^{a}$\\
 $^a$ Department of Physics, Carnegie-Mellon University,
Pittsburgh, PA 15213\\
 $^b$Department of Physics, North Carolina State University,
Raleigh, NC 27695-8202}
\date{\today}
\maketitle
\begin{abstract}
Using the light-front quark model, we calculate the transition form
factors, decay rates, and longitudinal lepton
polarization asymmetries for the exclusive rare
$B\to K\ell^+\ell^-$($\ell=e,\mu,\tau$) decays within the standard model.
Evaluating the timelike form factors,
we use the analytic continuation method in $q^+=0$ frame to obtain the
form factors $F_+$ and $F_T$, which are free from zero-mode.
The form factor $F_-$ which is not free from zero-mode in $q^+ = 0$ frame
and contaminated by the higher(or nonvalence) Fock states in $q^+ \neq 0$
frame is obtained from an effective treatment for handling the nonvalence
contribution based on the Bethe-Salpeter formalism.
The covariance(i.e. frame-independence)
of our model calculation is discussed. We obtain the branching ratios
for ${\rm BR}(B\to K\ell^+\ell^-)$ as
$4.96\times10^{-7}|V_{ts}/V_{cb}|^2$ for
$\ell=e,\mu$ and $1.27\times 10^{-7}|V_{ts}/V_{cb}|^2$ for
$\ell=\tau$.
\end{abstract}
\section{Introduction }
The upcoming and currently operating B factories BaBar at SLAC, Belle at
KEK, LHCB at CERN and B-TeV at Fermilab as well as the planned $\tau$-Charm
factory CLEO at Cornell make the precision test of standard
model(SM) and beyond SM ever more promising~\cite{Bfactory}.
Especially, a stringent test on the unitarity of
Cabibbo-Kobayashi-Maskawa (CKM) mixing matrix in SM
will be made by these facilities.
Accurate analyses of exclusive
semileptonic B-decays as well as rare B-decays are thus
strongly demanded for such precision tests.
One of the physics programs at the B factories is the exclusive rare
B decays induced by the flavor-changing neutral current(FCNC) transition.
Since in the standard model they are forbidden at tree level
and occur at the lowest order only through one-loop (Penguin)
diagrams~\cite{GWS,BM,Misiak,TI,AMM}, the rare B decays are well suited
to test the SM and search for physics beyond the SM.
While the experimental tests of exclusive decays are much easier than
those of inclusive ones, the theoretical understanding of exlcusive
decays is complicated mainly due to the nonperturbative hadronic
form factors entered in the long distance nonperturbative contributions.
The calculations of hadronic form factors for rare B decays have been
investigated by various theoretical approaches, such as
relativistic quark model~\cite{JW,GK,Mel1,MN}, heavy quark theory~\cite{HQ},
three point QCD sum rules~\cite{Col}, light cone QCD
sum rule~\cite{PB,BB,AKOS,ABHH}, and chiral perturbation theory~\cite{Cas,Du}.
Perhaps, one of the most well-suited formulations for the analysis of exclusive
processes involving hadrons may be provided in the framework of
light-front quantization~\cite{BPP}.

The aim of the present work is to calculate the hadronic form factors,
decay rates and the longitudinal lepton polarization asymmetries
for $B\to K\ell^+\ell^-$($\ell=e,\mu$, and $\tau$) decays
within the framework of the SM, using our light-front constituent
quark model(LFCQM or simply LFQM))~\cite{CJ1,CJ_PLB1,JC_E,CJK} based
on the LF quantization.
The longitudinal lepton polarization, as another parity-violating
observable, is an important asymmetry~\cite{Hew} and could be
measured by the above mentioned B factories.
In particular, the $\tau$ channel would be more accessible
experimentally than $e$- or $\mu$-channels
since the lepton polarization asymmetries
in the SM are known to be proportional to the lepton
mass. Although some recent works~\cite{AC} have studied the lepton
polarizations using the general form of the effective Hamiltonian
including all possible forms of interactions, we shall analyze
them within the SM as many others did.

Our LFQM~\cite{CJ1,CJ_PLB1,JC_E,CJK} used in the
present analysis has several salient features compared to
other LFQM~\cite{JW,GK} analysis: (1) We have implemented the variational
principle to the QCD motivated effective LF Hamiltonian to enable
us to analyze the meson mass spectra as well as various wavefunction-related
observables such as decay constants, electromagnetic form factors of mesons
in spacelike ($q^2<0$) region~\cite{CJ1}.
(2) We have performed the analytical continuation of the weak form
factors from spacelike region to the entire (physical) timelike region
to obtain the weak form factors for the exclusive semileptonic decays
of pseudoscalar mesons~\cite{CJ_PLB1}.
(3) We have recently presented in~\cite{JC_E} an effective treatment of
handling the higher Fock state (or nonvalence) contribution to the weak
form factor in $q^+>0$ frames, based on the Bethe-Salpeter(BS)
formalism (see also~\cite{CJK}).

The explicit demonstration
of our analytic continuation method using the exactly solvable model
of ($3+1$)-dimensional scalar field theory model can be found in~\cite{Anal}.
The Drell-Yan-West ($q^{+}$=$q^0$+$q^z$=0) frame is useful because
only valence contributions are needed as far as the ``$+$"-component
of the current is used.
Our analytic solution in the $q^+$=0 frame as a direct application
to the timelike region differs from the method used in~\cite{JW,GK} where
the authors used a simple parametric formula
extracted from the small $q^2$ behavior of a form factor.
However, some of the form factors in
timelike exclusive processes
receive higher Fock state contributions(i.e. zero-mode in $q^+=0$ frame
or nonvalence contribution in $q^+\neq 0$ frame)
within the framework of LF quantization. Thus, it is necessary
to include either zero-mode contribution(if working in $q^+=0$ frame)
or the nonvalence contribution (if working  in $q^+\neq 0$ frame) to
obtain such form factors.
Specifically, in the present analysis of exclusive rare
$B\to K\ell^+\ell^-$ decays, three independent hadronic form factors, i.e.
$F_+(q^2)$, $F_-(q^2)$ from the $V$-$A$(vector-axial vector) current,
and $F_T(q^2)$ from the tensor
current, are needed. While the two form factors $F_+$ and $F_T$
can be obtained from only valence contribution in $q^+=0$ frame
without encountering the zero-mode complication~\cite{Zero},
it is necessary to include the nonvalence
contribution for the calculation of the form factor $F_-$.
Our effective method\cite{JC_E} of calculating
novalence contributions has been shown to be quite reliable by checking
the covariance of the model.
Thus, we utilize both the analytic method in $q^+=0$ frame to obtain
($F_+,F_T$) and the effective method in $q^+>0$ frame to obtain $F_-$,
respectively.

The paper is organized as follows. In Sec. II, we
discuss the standard model effective Hamiltonian for the
exclusive rare $B\to K\ell^+\ell^-$ decays and reproduce the
QCD Wilson coefficients necessary in our analysis.
The formulas of the hadronic form factors, differential decay rates,
and the longitudinal lepton polarization asymmetries
are also introduced in this section.
In Sec. III, we calculate the weak form factors
$F_+(q^2),F_-(q^2)$ and $F_T(q^2)$ using our LFQM.
To obtain $F_+(q^2)$ and $F_T(q^2)$, we use the $q^{+}=0$
frame (i.e. $q^2=-\vec{q}^2_\perp<0$) and then analytically
continue the results to the timelike $q^{2}>0$
region by changing $q_{\perp}$ to $iq_{\perp}$ in the
form factors. The form factor $F_-(q^2)$ is obtained from our
effective method~\cite{JC_E} in purely longitudinal $q^+>0$ frames
(i.e. $q^2=q^+q^->0$).
 In Sec. IV, our numerical results, i.e. the form factors, decay rates,
and the longitudinal lepton polarization asymmetries
for $B\to K\ell^+\ell^-$ decays, are presented
and compared with the experimental data as well as other theoretical
results. Summary and discussion of our main results follow in Sec. V.
In the Appendix A, we list the QCD Wilson coefficients
necessary for the rare $B\to K$ transition.
In the Appendix B, we show the derivation of the differential
decay rate for $B\to K\ell^+\ell^-$ in the case of nonzero
lepton($m_\ell\neq 0$) mass.
In Appendix C, we show the generic form of our analytic solutions
for the weak form factors in timelike region.

\section{Overview of Effective Hamiltonian in Operator Basis}
The rare $b\to s\ell^+\ell^-$ decay process can be represented
in terms of the Wilson coefficients of the effective Hamiltonian
obtained after integrating out the heavy top quark and the $W^{\pm}$
bosons~\cite{GWS}, i.e.
\be\label{Hamil}
{\cal H}_{\rm eff}=\frac{4G_F}{\sqrt{2}}V_{tb}V^*_{ts}
\sum_i C_i(\tilde\mu)O_i(\tilde\mu),
\ee
where $G_F$ is the Fermi constant, $V_{ij}$ are the
CKM matrix elements and $C_i(\tilde\mu)$ are the Wilson coefficients.
It is known that the Wilson coefficients $C_3-C_6$ of QCD penguin
operators $O_3-O_6$ are small enough to be neglected and also the
operator $O_{8}$($\sim G^{a}_{\mu\nu}$, strong interaction field
strength tensor) does not contribute
to $b\to s\ell^+\ell^-$ transition.  Thus, the relevant basis
operators $O_i(\tilde\mu)$ to the rare $b\to s\ell^+\ell^-$ decay are
\bea\label{OPE}
O_1&=& (\bar{s}_\alpha \gamma^\mu P_L b_\alpha)
(\bar{c}_\beta\gamma^\mu P_L c_\beta),\nonumber\\
O_2&=& (\bar{s}_\alpha \gamma^\mu P_L b_\beta)
(\bar{c}_\beta\gamma^\mu P_L c_\alpha),\nonumber\\
O_7&=&\frac{e}{16\pi^2}m_b(\bar{s}_\alpha\sigma_{\mu\nu}
P_R b_\alpha)F^{\mu\nu},\nonumber\\
O_9&=&\frac{e^2}{16\pi^2}(\bar{s}_\alpha\gamma^\mu P_L b_\alpha)
(\bar{\ell}\gamma_\mu\ell),\nonumber\\
O_{10}&=& \frac{e^2}{16\pi^2}(\bar{s}_\alpha\gamma^\mu P_L b_\alpha)
(\bar{\ell}\gamma_\mu\gamma_5\ell),
\eea
where $P_{L(R)}=(1\mp\gamma_5)/2$ is the chiral projection operator and
$F^{\mu\nu}$ is the electromagnetic interaction field strength tensor.
The Lorentz and color indices are denoted as
$\mu$(and $\nu$) and $\alpha$(and $\beta$), respectively.
The renormalization scale $\tilde\mu$ in Eq.~(\ref{Hamil}) is usually
chosen to be $\tilde\mu\simeq m_b$ in order to avoid large logarithms,
ln($M_W/m_b$), in the matrix elements of the operators $O_i$.
The Wilson coefficients $C_i(m_b)$ determined by the renormalization
group equations(RGE) from the perturbative values $C_i(M_W)$ are
given in the literature(see, for example~\cite{BM,Misiak}).

Since the operators $O_1$ and $O_2$ contribute to $b\to s\ell^+\ell^-$
through $c\bar{c}$-loops which again couple to $\ell^+\ell^-$ through
virtual photon, they can be incorporated into
an ``effective" $O_9$. The resulting effective
Hamiltonian in Eq.~(\ref{Hamil}) has the following
structure(neglecting the strange quark mass)
\bea\label{Hll}
{\cal H}^{\ell^+\ell^-}_{\rm eff}&=&
\frac{4G_F}{\sqrt{2}}\frac{e^2}{16\pi^2}V^*_{ts}V_{tb}
\biggl[-\frac{2iC_7(m_b)m_b}{q^2}\bar{s}\sigma_{\mu\nu}q^\nu P_R b
\bar{\ell}\gamma^\mu\ell \nonumber\\
&+& C^{\rm eff}_9(m_b)\bar{s}\gamma_\mu P_L b\bar{\ell}\gamma^\mu\ell
+ C_{10}(m_b)\bar{s}\gamma_\mu P_L b\bar{\ell}\gamma^\mu\gamma_5\ell
\biggr].\nonumber\\
\eea

The effective Wilson coefficient
$C^{\rm eff}_9$($\hat{s}$=$q^2/m^2_b$) is given
by~\cite{AMM,KMS,AKS}
\bea\label{C9}
C^{\rm eff}_9(\hat{s})&\equiv&
\tilde{C}^{\rm eff}_9(\hat{s}) + Y_{\rm LD}(\hat{s}),\nonumber\\
&=&C_9\biggl(1+\frac{\alpha_s(\mu)}{\pi}\omega(\hat{s})\biggr)
+Y_{\rm SD}(\hat{s})+Y_{\rm LD}(\hat{s}),
\eea
where the function $Y_{\rm SD}(\hat{s})$ is the one-loop
matrix element of $O_9$, $Y_{\rm LD}(\hat{s})$ describes
the long distance contributions due to the charmonium vector
$J/\psi,\psi',\cdots$ resonances via
$B\to K(J/\psi,\psi',\cdots)\to K\ell^+\ell^-$,
and $\omega(\hat{s})$
represents the one-gluon correction to the matrix element
of $O_9$. Their explicit forms are given in the
literature~\cite{BM,Misiak,KMS,AKS,LW} and also in the Appendix A of this
work. For the numerical values of the Wilson coefficients and relevant
parameters in obtaining Eq.~(\ref{C9}), we
use the results given by Refs.~\cite{AKS,LW}: $m_t=175$ GeV, $m_b=4.8$
GeV, $m_c=1.4$ GeV, $\alpha_s(M_W)=0.12$, $\alpha_s(m_b)=0.22$,
$C_1=-0.26$, $C_2=1.11$, $C_3=0.01$, $C_4=-0.03$, $C_5=0.008$,
$C_6=-0.03$, $C_7=-0.32$, $C_9=4.26$, and
$C_{10}=-4.62$.

\begin{figure}
\centerline{\psfig{figure=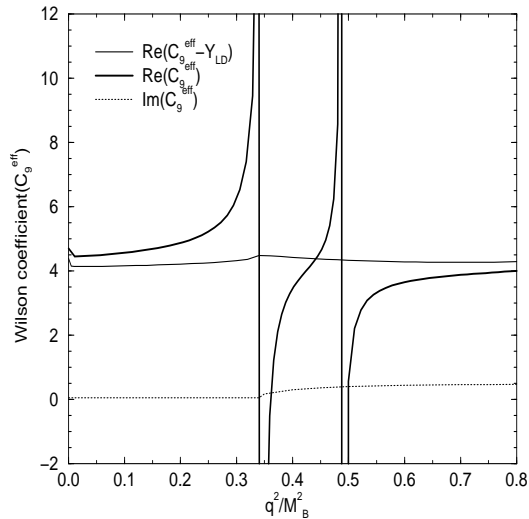,height=8cm,width=8cm}}
\caption{ The effective Wilson coefficient $C^{\rm eff}_9$
as a function of $\hat{s}=q^2/M^2_B$. As the real part of
$C^{\rm eff}_9$, the thick(thin) solid line represents the
results with(without) LD contribution, i.e.
$C^{\rm eff}_9(\tilde{C}^{\rm eff}_9)$.
The imaginary (dotted line) part of $C^{\rm eff}_9$ is the
result without LD contribution.
\label{fig_C9}}
\end{figure}

In  Fig.~\ref{fig_C9}, we plot the effective
Wilson coefficient $C^{\rm eff}_9$ as a function of $\hat{s}$.
As the real part of $C^{\rm eff}_9$, the thick(thin) solid line
represents the result with(without) LD contribution, i.e.
${\rm Re}(C^{\rm eff}_9)({\rm Re}(\tilde{C}^{\rm eff}_9))$.
The imaginary (dotted line) part of $C^{\rm eff}_9$ is the result
without LD contribution, ${\rm Im}(\tilde{C}^{\rm eff}_9)$.
In our numerical calculation of $C^{\rm eff}_9$(thick solid
lines), we include two charmonium
vector $J/\psi(1S)$ and $\psi'(2S)$ resonances(see Appendix A).
The cusp of Re($\tilde{C}^{\rm eff}_9$) at
$\hat{s}=4(m_c/m_b)^2\simeq 0.34$ as shown in
Fig.~\ref{fig_C9} (thin line) is due to the $c\bar{c}$-loop
contribution from $Y_{SD}(\hat{s})$[see Eqs.~(\ref{YSD}) and
(\ref{gms}) in Appendix A].
In Fig.~\ref{fig_C9}, one can also find that
${\rm Re}(\tilde{C}^{\rm eff}_9)\gg {\rm Im}(\tilde{C}^{\rm eff}_9)$.

The long-distance contribution to the exclusive $B\to K$ decay is
contained in the meson matrix elements of the bilinear quark currents
appearing in ${\cal H}^{\ell^+\ell^-}_{\rm eff}$ given by Eq.~(\ref{Hll}).
The matrix elements of the hadronic currents for $B\to K$ transition
can be parametrized in terms of hadronic form factors as follows
\bea\label{Jmu}
J^\mu&\equiv&\la K|\bar{s}\gamma^{\mu}P_L b|B\ra=
\frac{1}{2}[F_{+}(q^{2})P^\mu + F_{-}(q^{2})q^\mu],
\eea
and
\bea\label{JTmu}
J^\mu_T&\equiv&\la K|\bar{s}i\sigma^{\mu\nu}q_\nu P_R b|B\ra
\nonumber\\
&=& \frac{1}{2(M_B+M_K)}[q^2 P^\mu - (M^2_B-M^2_K)q^\mu] F_T(q^2),
\eea
where $P=P_B+P_K$ and $q=P_B-P_K$ is the four-momentum
transfer to the lepton pair and $4m^{2}_{l}\leq q^{2}\leq(M_B-M_K)^2$.
We use the convention $\sigma^{\mu\nu}=(i/2)[\gamma^\mu,\gamma^\nu]$ for
the antisymmetric tensor.
Sometimes it is useful to express Eq.~(\ref{Jmu}) in terms
of $F_+(q^2)$ and $F_0(q^2)$, which are related to the exchange
of $1^-$ and $0^+$, respectively, and satisfy the following relations:
\be\label{F0}
F_+(0)=F_0(0),\;
F_0(q^2)=F_+(q^2) + \frac{q^2}{M^2_B-M^2_K}F_-(q^2).
\ee
With the help of the effective Hamiltonian in Eq.~(\ref{Hll}) and
Eqs.~(\ref{Jmu}) and~(\ref{JTmu}),  the transition amplitude
for the $B\to K\ell^+\ell^-$ decay can be written as
\bea\label{TranA}
{\cal M} &=& \la K\ell^+\ell^-|{\cal H}_{\rm eff}|B\ra \nonumber\\
&=&\frac{4G_F}{\sqrt{2}}\frac{\alpha}{4\pi}V^*_{ts}V_{tb}
\biggl\{\biggl[C^{\rm eff}_9 J_\mu - \frac{2m_b}{q^2}C_7 J^T_\mu\biggr]
\bar{\ell}\gamma^\mu\ell \nonumber\\
&&\hspace{2.5cm}
+ C_{10}J_\mu \bar{\ell}\gamma^\mu\gamma_5\ell \biggr\},
\eea
where $\alpha=e^2/4\pi$ is the fine structure constant.
The differential decay rate for the exlcusive
rare $B\to K\ell^+\ell^-$ with nonzero lepton mass($m_\ell\neq 0$)
is given by~(see Appendix B for the detailed derivation)
\bea\label{DDR}
\frac{d\Gamma}{d\hat{s}}
&=&\frac{M^5_BG^2_F}{3\cdot2^9\pi^5}\alpha^2
|V^*_{ts}V_{tb}|^2\hat{\phi}^{1/2}
\biggl(1-4\frac{\hat{m}_\ell}{\hat{s}}\biggr)^{1/2}
\nonumber\\
&&\times
\biggl[\hat{\phi}\biggl(1+2\frac{\hat{m}_\ell}{\hat{s}}\biggr)
F_{T+} + 6\frac{\hat{m}_\ell}{\hat{s}}F_{0+} \biggr],
\eea
where
\bea\label{DDR2}
F_{T+}&=&|C^{\rm eff}_9 F_+ - \frac{2C_7}{1+\sqrt{\hat{r}}}F_T|^2
+ |C_{10}|^2|F_+|^2,\nonumber\\
F_{0+}&=& |C_{10}|^2 [(1-\hat{r})^2|F_0|^2-\hat{\phi}|F_+|^2],
\nonumber\\
\hat{\phi}&=& (\hat{s}-1-\hat{r})^2-4\hat{r},
\eea
with $\hat{s}=q^2/M^2_B$, $\hat{m}_\ell=m^2_\ell/M^2_B$, and
$\hat{r}=M^2_K/M^2_B$.
We used $m_b\simeq M_B$ in derivation of Eq.~(\ref{DDR}).
Note also from Eqs.~(\ref{DDR}) and~(\ref{DDR2}) that
the form factor $F_-(q^2)$(or $F_0(q^2)$) contributes
only in the nonzero lepton($m_\ell\neq 0$) mass limit.
Dividing Eq.~(\ref{DDR}) by the total width
of the $B$ meson, which is estimated to be~\cite{JW,DT}
\be\label{TD}
\Gamma_{\rm tot}=\frac{f M^5_BG^2_F}{192\pi^3}|V_{cb}|^2,\;
f\simeq 3.0,
\ee
one can obtain the differential branching ratio
$dBR(B\to K\ell^+\ell^-)/d\hat{s}
=(d\Gamma(B\to K\ell^+\ell^-)/\Gamma_{\rm tot})/d\hat{s}$
\footnote{With $f=3$ and
the central value of $|V_{cb}|=0.0402$~\cite{PDG}, we obtain
$\tau_B\simeq 1.688$ ps while $\tau^{\rm exp}_{B^{\pm}}=(1.653\pm0.028)$ ps.
Since our numerical results of the branching ratios are obtained
from using Eq.~(\ref{TD}), approximately 2$\%$ theoretical error
due to the lifetime of $B$ meson is understood.}.


As another interesting observable, the longitudinal lepton
polarization asymmetry(LPA), is defined as
\be\label{LPA}
P_L(\hat{s})=\frac{d\Gamma_{h=-1}/d\hat{s}-d\Gamma_{h=1}/d\hat{s}}
{d\Gamma_{h=-1}/d\hat{s} +d\Gamma_{h=1}/d\hat{s}},
\ee
where $h=+1(-1)$ denotes right (left) handed $\ell^-$ in the final state.
From Eq.~(\ref{DDR}), one obtains for $B\to K\ell^+\ell^-$
\be\label{LPA_BK}
P_L(\hat{s})=\frac{
2\biggl(1-4\frac{\hat{m}_\ell}{\hat{s}}\biggr)^{1/2}
\hat{\phi}C_{10}F_{+}
\biggl[F_+ {\rm Re}C^{\rm eff}_9 -
\frac{2C_7}{1+\sqrt{\hat{r}}}F_T\biggr] }
{ \biggl[\hat{\phi}\biggl(1+2\frac{\hat{m}_\ell}{\hat{s}}\biggr)
F_{T+} + 6\frac{\hat{m}_\ell}{\hat{s}}F_{0+} \biggr] }.
\ee
Note that our formulas for the differential decay rate
in Eq.~(\ref{DDR}) and the LPA in Eq.~(\ref{LPA_BK}) are written in
terms of ($F_+,F_0,F_T$) instead of ($F_+,F_-,F_T$) as obtained
in Refs.~\cite{GK,MN}. However, our formulas and those in~\cite{GK,MN}
are equivalent with each other once we rearrange our formulas in terms
of ($F_+,F_-,F_T$). One nice feature of using $F_0$ in the decay rate
formula is to separate the $F_0$ contribution from the total rate as
we shall show later.

\section{ Form Factor Calculation in Light-Front Quark Model}

\subsection{Analytic calculation in $q^{+}=0$ frame}
As shown in Eq.~(\ref{DDR}), only two weak form factors
$F_+(q^2)$ and $F_T(q^2)$ are necessary for the
massless($m_\ell=0$) rare exclusive semileptonic $b\to s\ell^+\ell^-$
process. The form factors $F_+(q^2)$ and $F_T(q^2)$
can be obtained in $q^+=0$ frame with the ``good" component of
currents, i.e. $\mu=+$, without encountering zero-mode
contributions~\cite{Zero}.
Thus, we shall perform our light-front quark
model calculation in the $q^+=0$ frame,
where $q^2=q^+q^- -{\vec{q}_\perp}^2=-\vec{q}^2_\perp<0$, and then
analytically continue the form factors
$F_i(\vec{q}^2_\perp)(i=+,T)$ in spacelike region to the
timelike $q^2>0$ region by changing $\vec{q}_\perp$ to $i\vec{q}_\perp$
in the form factor.

The quark momentum variables for $P_B(q_{1}\bar{q})\to
P_K(q_{2}\bar{q})$ transitions in the $q^{+}=0$ frame are given by
\bea\label{vari}
p^{+}_{1}&=&(1-x)P^{+}_{1},\hspace{1.5cm}
p^{+}_{\bar{q}}=xP^{+}_{1},\nonumber\\
{\vec p}_{1\perp}&=&(1-x){\vec P}_{1\perp} + {\vec k}_{\perp},\hspace{.5cm}
{\vec p}_{\bar{q}\perp}= x{\vec P}_{1\perp} - {\vec k}_{\perp},
\nonumber\\
p^{+}_{2}&=&(1-x)P^{+}_{2},\hspace{1.5cm}
p'^{+}_{\bar{q}}=xP^{+}_{2},\nonumber\\
{\vec p}_{2\perp}&=&(1-x){\vec P}_{2\perp} +
{\vec k'}_{\perp},\hspace{.5cm}
{\vec p'}_{\bar{q}\perp}= x{\vec P}_{2\perp} - {\vec k'}_{\perp},
\eea
which require that $p^{+}_{\bar{q}}=p'^{+}_{\bar{q}}$
and ${\vec p}_{\bar{q}\perp}={\vec p'}_{\bar{q}\perp}$.
For $B\to K$ transitions, one has $m_1=m_b$, $m_2=m_s$, and
$m_{\bar{q}}=m_{u}$.
Our analysis for $b\to s\ell^+\ell^-$ decays will be carried out
in this $q^{+}=0$ frame and the decaying hadron (B-meson) is
at rest, i.e. ${\vec P}_{1\perp}=0$.

The matrix elements of the currents $J^\mu$ in Eq.~(\ref{Jmu})
and $J^\mu_T$ in Eq.~(\ref{JTmu}) are
obtained by the convolution formula of the initial and final state
light-front wave functions as follows
\bea\label{Con}
\la P_2|\bar{q}_2\Gamma^\mu q_1|P_1\ra
&=&\sum_{{\lambda}'s}\int d^3\vec{p}_{\bar{q}}\;
\phi_{2}(x,{\vec k'}_\perp)\phi_{1}(x, {\vec k}_\perp)
\nonumber\\
&&\times
{\cal R}^{00^\dagger}_{\lambda_2\bar{\lambda}}
\frac{\bar{u}_{\lambda_2}(p_2)}{\sqrt{p^+_2}}\Gamma^\mu
\frac{u_{\lambda_1}(p_1)}{\sqrt{p^+_1}}
{\cal R}^{00}_{\lambda_1\bar{\lambda}},
\eea
where $\Gamma^\mu=\gamma^\mu P_L$ for $J^\mu$ in Eq.~(\ref{Jmu})
and $i\sigma^{\mu\nu}q_\nu P_R$ for $J^{\mu}_T$ in Eq.~(\ref{JTmu}),
respectively.
The measure $[d^3\vec{p}_{\bar{q}}]$ in Eq.~(\ref{Con})
is written in terms of light-front variables as
\bea\label{d3p}
d^3\vec{p}_{\bar{q}}&=&P^+_1dxd^2\vec{k}_\perp
\sqrt{\frac{\partial k'_z}{\partial x}}
\sqrt{\frac{\partial k_z}{\partial x}},
\eea
where ${\partial k_{z}}/{\partial x}$ is the Jacobian of
the variable transformation $\{x,{\vec k}_{\perp}\}\to {\vec k}=
(k_{z},{\vec k}_{\perp})$ defined by
\bea\label{jacob}
\frac{\partial k_z}{\partial x}&=& \frac{M_{0}}{4x(1-x)}
\biggl[1-\biggl(\frac{m^2_q-m^2_{\bar{q}}}{M^2_{0}}\biggr)^2
\biggr],\\
M^2_0&=& \frac{m^2_q+\vec{k}^2_\perp}{1-x}
         + \frac{m^2_{\bar q}+\vec{k}^2_\perp}{x}.
\eea
The spin-orbit wave function
${\cal R}^{JJ_{z}}_{\lambda_{q},\lambda_{\bar{q}}}(x,{\vec k}_{\perp})$
is obtained by the interaction-independent Melosh transformation.
The explicit covariant form for a pseudoscalar($J=0,J_z=0$) meson is
given by
\be\label{SO}
{\cal R}^{J=0,J_{z}=0}_{\lambda_{q},\lambda_{\bar{q}}}(x,{\vec k}_{\perp})
=\frac{\bar{u}(p_q,\lambda_q)\gamma^5 v(p_{\bar q},\lambda_{\bar q})}
{\sqrt{2}\sqrt{M^2_0-(m^2_q-m^2_{\bar q})^2}},
\ee
where ${\lambda}'s$ are light-front helicities.
Our radial wave function is given by the gaussian trial function for the
variational principle to the QCD-motivated effective light-front
Hamiltonian~\cite{CJ1}:
\be\label{Rad}
\phi(x,{\vec k}_{\perp})=
\biggl(\frac{1}{\pi^{3/2}\beta^{3}}\biggr)^{1/2}
\exp(-\vec{k}^{2}/2\beta^{2}),
\ee
which is normalized as $\int d^3k|\phi(x,{\vec k}_{\perp})|^2=1$,
where $\vec{k}^2=\vec{k}^2_\perp + k^2_z$ and $k_z$ is given by
\be\label{kz}
k_z=(x-\frac{1}{2})M_0 + \frac{m^2_q-m^2_{\bar q}}{2M_0}.
\ee

Then, the sum of the light-front spinors over the helicities
in Eq.~(\ref{Con}) are obtained as
\bea\label{spinor}
&&\sum_{{\lambda}'s}v_{\lambda_{\bar q}}^\dagger(p_{\bar q})
\gamma^5\bar{u}_{\lambda_2}^\dagger(p_2)
\bar{u}_{\lambda_2}(p_2)\Gamma^\mu
u_{\lambda_1}(p_1)\bar{u}_{\lambda_1}(p_1)\gamma^5
v_{\lambda_{\bar q}}(p_{\bar q})\nonumber\\
&&= {\rm Tr}\biggl[(\not\!p_{\bar q}-m_{\bar q})\gamma^5
(\not\!p_2+m_2)\Gamma^\mu(\not\!p_1+m_1)\gamma^5\biggr].
\eea
Using the matrix element of the ``$+$" component of the
currents($\mu=+$),
and the particle on-mass shell condition, i.e. the light-front energy
$p^-_i=(\vec{p}^2_{i\perp}+m^2_i)/p^+_i$($i=1,2$ and $\bar{q}$)
in Eq.~(\ref{spinor}),
we obtain the weak form factors $F_{+}(\vec{q}_{\perp}^{2})$ and
$F_T(\vec{q}_{\perp}^{2})$ as follows
\bea\label{FP}
F_{+}(\vec{q}_{\perp}^{2}) &=&
\int^{1}_{0}dx\int d^{2}{\vec k}_{\perp}
\sqrt{\frac{\partial k'_z}{\partial x}}
\sqrt{\frac{\partial k_z}{\partial x}}
\phi_{2}(x,{\vec k'}_{\perp})\phi_{1}(x,{\vec k}_{\perp})
\nonumber\\
& &\times
\frac{A_{1} A_{2}+{\vec k}_{\perp}\cdot{\vec k'}_{\perp}}
{ \sqrt{ A_{1}^{2}+ \vec{k}^{2}_{\perp}}\sqrt{ A_{2}^{2}+
\vec{k'}^{2}_{\perp}} },
\end{eqnarray}
and
\bea\label{FT}
F_{T}(\vec{q}_{\perp}^{2}) &=&\hspace{-0.1cm}
-\hspace{-0.1cm}\int^{1}_{0}dx\int d^{2}{\vec k}_{\perp}
\sqrt{\frac{\partial k'_z}{\partial x}}
\sqrt{\frac{\partial k_z}{\partial x}}
\phi_{2}(x,{\vec k'}_{\perp})\phi_{1}(x,{\vec k}_{\perp})
\nonumber\\
&& \times
\frac{M_B+M_K}{(1-x)\tilde{M}_{0}\tilde{M'}_{0}}
\biggl[(m_2-m_1)\frac{\vec{k}_\perp\cdot\vec{q}_\perp}{\vec{q}^2_\perp}
+ A_1\biggr],\nonumber\\
\eea
where $A_{i}=xm_{i} + (1-x)m_{\bar{q}}(i=1,2)$,
$\tilde{M}_0=\sqrt{M^2_0-(m_q-m_{\bar q})^2}$,
and ${\vec k'}_{\perp}={\vec k}_{\perp}-x{\vec q}_{\perp}$.
The primed factors in Eqs.~(\ref{FP}) and~(\ref{FT}) are the
functions of final state momenta, e.g. $k'_z=k'_z(x,\vec{k'}_\perp)$
and $\tilde{M'}_0=\tilde{M'}_0(x,\vec{k'}_\perp)$.
Since the weak form factors $F_{+}(\vec{q}^2_\perp)$ in Eq.~(\ref{FP})
and $F_{T}({\vec q_\perp}^2)$ in Eq.~(\ref{FT}) are defined in
the spacelike($q^2<0$) region, we then analytically continue them to
the timelike $q^{2}>0$ region by replacing $q_{\perp}$ with $iq_{\perp}$ in
the form factors.
We describe in Appendix C our procedure of analytic continuation of
the weak form factors.

Our analytic solutions will be compared with the following parametric
form used by many others~\cite{JW,GK,Mel1,PB,AKS}
\be\label{Pole}
F(q^2)=\frac{F(0)}{1-\sigma_1 q^2+\sigma_2 q^4},
\ee
where the parameters $\sigma_1$ and $\sigma_2$ are determined by
the first and second derivatives of $F(q^2)$ at $q^2=0$.

\subsection{Effective calculation in $q^+ > 0$ frame}

Our effective calculation of weak form factors is performed
in the purely longitudinal momentum frame~\cite{JC_E,Zero}
where $q^+>0$ and ${\vec P}_{1\perp}={\vec P}_{2\perp}=0$
so that the momentum transfer square $q^2=q^+q^->0$ is timelike.

\begin{figure}
\centerline{\psfig{figure=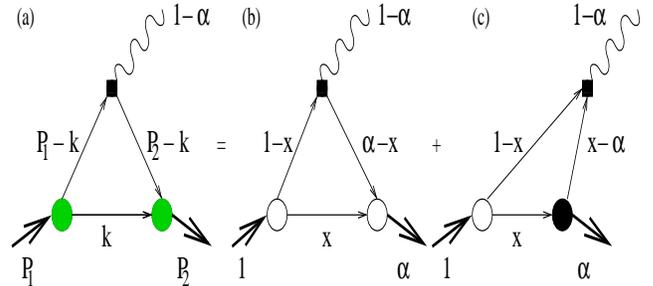,height=1.5in,width=3.3in}}
\caption{The covariant diagram (a) corresponds to the sum of the
LF valence diagram (b) defined in $0<x<\al$ region and the
nonvalence diagram (c) defined in $\al<x<1$ region. The large white
and black blobs at the meson-quark vertices in (b) and (c) represent
the ordinary LF wave function and the nonvalence wave function vertices,
respectively. The small black box at the quark-gauge boson vertex
indicates the insertion of the relevant Wilson operator. \label{fig_HF} }
\end{figure}

One can then easily obtain $q^2$ in terms of the momentum fraction
$\alpha=P^{+}_{2}/P^{+}_{1}=1-q^{+}/P^{+}_{1}$
as $q^{2}=(1-\alpha)(M^{2}_{1}-M^{2}_{2}/\al)$.
Accordingly, the two solutions for $\alpha$ are given by
\be\label{apm}
\alpha_{\pm}=\frac{M_{2}}{M_{1}}\biggl[
\frac{ M^{2}_{1}+M^{2}_{2}-q^{2}}{2M_{1}M_{2}}
\pm \sqrt{\biggl(\frac{ M^{2}_{1}+M^{2}_{2}-q^{2}}
{2M_{1}M_{2}}\biggr)^{2}-1} \biggr].
\ee
The $+(-)$ sign in Eq.~(\ref{apm}) corresponds to the daughter meson
recoiling in the positive(negative) $z$-direction relative to
the parent meson. At zero recoil($q^{2}=q^{2}_{\rm max}$) and
maximum recoil($q^{2}=0$), $\alpha_{\pm}$ are given by
\begin{eqnarray}\label{alimit}
&&\alpha_{+}(q^{2}_{\rm max})=
\alpha_{-}(q^{2}_{\rm max})=\frac{M_{2}}{M_{1}},
\nonumber\\
&&\alpha_{+}(0)=1,\hspace{0.5cm}
\alpha_{-}(0)=\biggl(\frac{M_{2}}{M_{1}}\biggr)^{2}.
\end{eqnarray}
The quark momentum variables in the $q^{+}>0$ frame are similar to
Eq.~(\ref{vari}) in the $q^+=0$ frame but the momentum transfer $q^2$
in $q^{+}>0$ frames flows through only longitudinal component of
quark and antiquark momenta, i.e.
\bea\label{vari_nq}
p^{+}_{1}&=&(1-x)P^{+}_{1},\;
p^{+}_{\bar{q}}=xP^{+}_{1},\;
{\vec p}_{1\perp}=-{\vec p}_{\bar{q}\perp}={\vec k}_{\perp},
\nonumber\\
p^{+}_{2}&=&(1-x')P^{+}_{2},\;
p'^{+}_{\bar{q}}=x'P^{+}_{2},\;
{\vec p}_{2\perp}= -{\vec p'}_{\bar{q}\perp}={\vec k}_{\perp},
\eea
where $x'=x/\al$ and ${\vec P}_{1\perp}={\vec P}_{2\perp}=0$ has been
used (see Fig.~\ref{fig_HF}).

The $\alpha_{\pm}$-independent form factors $F_{\pm}(q^{2})$
defined in $q^+>0$ frames are then obtained as follows
\be\label{fpm}
F_{\pm}(q^{2})=\pm \frac{(1\mp \alpha_{-})j^{+}(\alpha_{+}) -
(1\mp \alpha_{+})j^{+}(\alpha_{-})}
{\alpha_{+}-\alpha_{-}},
\ee
where $j^+(\al_{\pm})=\la K|\bar{s}\gamma^{+}P_L b|B\ra|_{\al_{\pm}}/P^+_1$
from Eq.~(\ref{Jmu}).

As shown in Fig.~\ref{fig_HF}, the $q^+$$>$0 frame requires not only
the particle-number-conserving (valence) Fock state contribution
in Fig.~\ref{fig_HF}(b) but also the particle-number-nonconserving
(nonvalence) Fock state contribution in Fig.~\ref{fig_HF}(c);
i.e.  $j^+(\al_{\pm})=j^+_{val}(\al_{\pm}) + j^+_{nv}(\al_{\pm})$
in Eq.~(\ref{fpm}). In our previous works~\cite{JC_E,CJK},
we have developed a new effective treatment  of the non-wave-function
vertex(black blob in Fig.~\ref{fig_HF}(c)) in the nonvalence diagram
arising from the quark-antiquark pair creation/annihilation.
Since the detailed procedures for obtaining the effective solution
for the non-wave-function vertex have been given in~\cite{JC_E,CJK},
here we briefly present the salient points of our effective
method~\cite{JC_E,CJK} and the final forms of the current matrix
elements for both valence and nonvalence diagrams.

\begin{figure}
\centerline{\psfig{figure=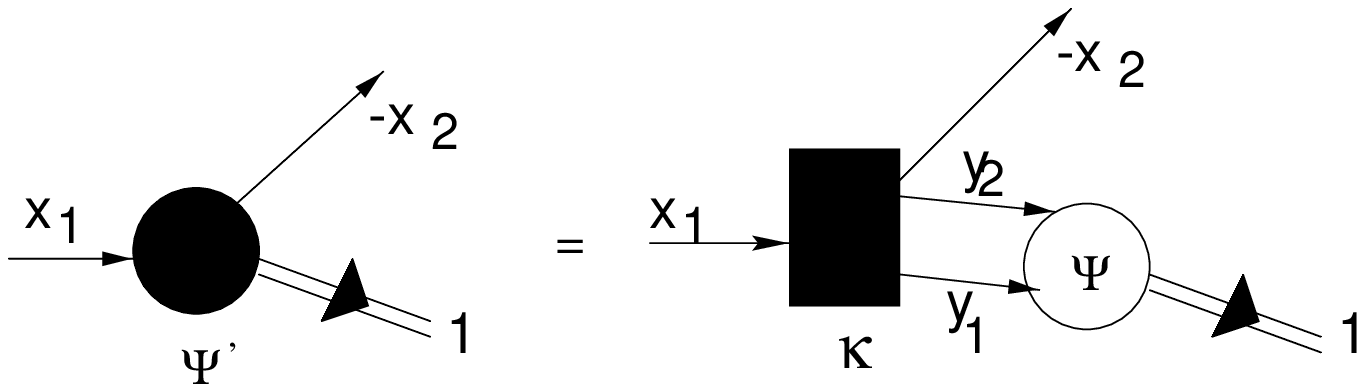,height=1.0in,width=3.2in}}
\caption{Non-wave-function vertex(black blob) linked to an ordinary
LF wave function(white blob).}\label{fig_SD}
\end{figure}

The essential feature of our approach is to consider the light-front
wave function as the solution of light-front
Bethe-Salpeter equation(LFBSE) given by
\bea\label{eq:SD}
&&(M^2-{\cal M}^{2}_0)\Psi(x_i,{\vec k}_{i\perp})
\nonumber\\
&&\;\; =\int[dy][d^2{\vec l}_{\perp}]
{\cal K}(x_i,{\vec k}_{i\perp};y_j,{\vec l}_{j\perp})
\Psi(y_j,{\vec l}_{j\perp}),
\eea
where ${\cal K}$ is the B-S kernel which in principle includes all the
higher Fock-state contributions,
${\cal M}^{2}_0=(m^2_1+{\vec k}^2_{1\perp})/x_1 +
(m^2_2+{\vec k}^2_{2\perp})/x_2$, and $\Psi(x_i,{\vec k}_{i\perp})$ is
the B-S amplitude.
Both the valence(white blob) and nonvalence(black blob) B-S amplitudes
are solutions to Eq.~(\ref{eq:SD}).
For the normal(or valence) B-S
amplitude, $x_1=x$ and $x_2=\al-x>0$, while for the nonvalence B-S
amplitude, $x_1=x$ and $x_2=\al-x<0$.
As illustrated in  Figs.~\ref{fig_HF}(b) and~(c), the nonvalence
B-S amplitude is an analytic continuation of the valence B-S
amplitude. In the LFQM the relationship between the B-S amplitudes
in the two regions is given by~\cite{JC_E,CJK}
\bea\label{eq:SD2}
&&(M^2-{\cal M}^{2}_0)\Psi'(x_i,{\vec k}_{i\perp})
\nonumber\\
&&\;\;=\int[dy][d^2{\vec l}_{\perp}]
{\cal K}(x_i,{\vec k}_{i\perp};y_j,{\vec l}_{j\perp})
\Psi(y_j,{\vec l}_{j\perp}),
\eea
where $\Psi'(x_i,{\vec k}_{i\perp})$ represents the nonvalence B-S amplitude
and again the kernel includes in principle all the higher Fock state
contributions because all the higher Fock components of the bound-state
are ultimately related to the lowest Fock component with the use of
the kernel. This is illustrated in Fig.~\ref{fig_SD}.

Equations~(\ref{eq:SD}) and~(\ref{eq:SD2}) are integral equations for
which one needs nonperturbative QCD to obtain the kernel. We do not
solve for the B-S amplitudes in this work, but a nice feature of
Eq.~(\ref{eq:SD2}) is a natural link between nonvalence B-S amplitude
$\Psi'$ and the valence one $\Psi$ which enables an application of a
light-front CQM even for the calculation of nonvalence contribution
in Fig.~\ref{fig_HF}(c).  In ($1+1$)-QCD models~\cite{Einhorn,Burk},
it is shown that expressions for the nonvalence vertex analogous
to our form given in Eq.~(\ref{eq:SD2}) are obtained.
With the iteration procedure given by Eq.~(\ref{eq:SD2})
in this $q^+>0$ frame, we obtain the current matrix element
of the nonvalence diagram in terms of light-front vertex function
and the gauge boson vertex function. The interested reader may consult
Refs.~\cite{JC_E,CJK} on this subject.

The matrix element of the valence current, $j^+_{val}$ in Eq.~(\ref{fpm}),
is given by
\bea{\label{eq:JV}}
j^{+}_{val}&=& \int^{\alpha}_{0}dx\int d^{2}{\vec k}_{\perp}
\sqrt{\frac{\partial k'_z}{\partial x'}}
\sqrt{\frac{\partial k_z}{\partial x}}
\phi_2(x',{\vec k}_\perp)\phi_1(x,{\vec k}_\perp)
\nonumber\\
& &\times
\frac{B_{1}B_{2}+{\vec k}^2_{\perp}}
{ \sqrt{ B_{1}^{2}+ \vec{k}^{2}_{\perp}}\sqrt{ B_{2}^{2}+
\vec{k}^{2}_{\perp}} },
\eea
where
\be\label{B1B2}
B_1=x  m_1 + (1-x) m_{\bar{q}},\;
B_2=x'  m_2 + (1-x') m_{\bar{q}},
\ee
and $k'_z=k_z(x',{\vec k}_\perp)$ in Eq.~(\ref{kz}).
The matrix element of the nonvalence current, $j^+_{nv}$
in Eq.~(\ref{fpm}), is obtained as
\bea{\label{eq:JNV}}
j^{+}_{nv}&=&\int^{1}_{\alpha}
\frac{dx}{x'(1-x')}
\int d^{2}{\vec k}_{\perp}
\sqrt{\frac{\partial k_z}{\partial x}}
\chi^g(x,{\vec k}_{\perp})\phi_1(x,{\vec k}_{\perp})
\nonumber\\
&&\times
\frac{ {\vec k}^2_\perp + B_1B_2 + x(1-x)(1-x')(M^2_1-M^2_0)}
{\sqrt{x(1-x)} \tilde{M}_{0}}
\nonumber\\
&&\times
\int \widehat{dy}\int d^2{\vec l}_{\perp}
\sqrt{\frac{\partial l_z}{\partial y}}
\frac{{\cal K}(x,{\vec k}_{\perp};y,{\vec l}_{\perp})}
{\tilde{M'}_{0}(y,{\vec{l}_\perp})}
\phi_{2}(y,{\vec l}_{\perp}),
\eea
where
\be\label{gaugeWF}
\chi^g(x,{\vec k}_\perp)=
\frac{1}{\alpha\biggl[\frac{q^2}{1-\alpha} -
\biggl(\frac{{\vec k}^2_\perp + m^2_1}{1-x}
+\frac{{\vec k}^2_\perp + m^2_2}{x -\alpha}\biggr)
\biggr]}
\ee
is the light-front vertex function of a gauge boson
\footnote{While one can in principle also consider the B-S
amplitude for $\chi^g$, we note that such extension does not
alter our results within our approximation in this work because both
hadron and gauge boson should share the same kernel.}
and $\widehat{dy}=dy/\sqrt{y(1-y)}$.
In derivation of Eq.~(\ref{eq:JNV}) with the ``+"-component
of the current, we also separate the on-mass
shell propagating part(i.e. the term proportional to
$({\vec k}^2_\perp + B_1B_2)$)
from the instantaneous part(i.e. the term proportional to
$x(1-x)(1-x')(M^2_1-M^2_0)$),
where the struck quarks ($m_1=m_b$ and $m_2=m_s$) are on-mass shell
and the spectator quark ($m_{\bar q}=m_u$) is off-mass shell.
Note that the instantaneous contribution exists only for the nonvalence
diagram as far as the ``$+$"-component of the current is used.
As we shall show in the next numerical section, the instantaneous
contribution to the weak form factors $F_{\pm}(q^2)$ for $B\to K$
transition is quite substantial near zero recoil.

Note that Eq.~(\ref{eq:SD2}) was used to obtain the last
term in Eq.~(\ref{eq:JNV}). While the relevant operator ${\cal K}$
is in general dependent on all internal momenta
$(x,{\vec k}_\perp; y,{\vec l}_\perp)$, the integral of ${\cal K}$
over $y$ and ${\vec l}_\perp$ in Eq.~(\ref{eq:JNV})
depends only on $x$ and ${\vec k}_\perp$, which we define
\be\label{GBK}
G_{BK}(x,{\vec k}_\perp)\equiv
\int \widehat{dy}\int d^2{\vec l}_{\perp}
\sqrt{\frac{\partial l_z}{\partial y}}
\frac{{\cal K}(x,{\vec k}_{\perp};y,{\vec l}_{\perp})}
{\tilde{M'}_{0}(y,{\vec{l}_\perp})}
\phi_{2}(y,{\vec l}_{\perp}).
\ee
In this work, we approximate $G_{BK}(x,{\vec k}_\perp)$ as a
constant which has been tested in our previous works~\cite{JC_E,CJK}
and proved to be a good approximation.
As we shall show in the next section, the reliability of this
approximation can be checked by examining the frame-independence of
our numerical results.

\section{Numerical results}
In our numerical calculation for the process of
$B\to K\ell^+\ell^-$ transition, we use the linear
potential parameters presented in Ref.~\cite{CJ_PLB1}.
Our predictions of the decay constants for $K$ and $B$
were reported~\cite{CJ1,CJ_PLB1} as $f_K$=161.4 MeV(Exp.=
159.8$\pm$1.4)~\cite{CJ1} and $f_B=171.4$ MeV~\cite{CJ_PLB1},
respectively.\footnote{The difference of decay constants between
this work and Refs.~\cite{CJ1,CJ_PLB1} is only due to the
definition, i.e. we use the definition
$\la 0|\bar{q}_2\gamma^\mu\gamma_5q_1|P\ra=if_PP^\mu$ in this work
so that $f^{\rm Exp.}_\pi=130.7\pm 0.1$ MeV while we used
$\la 0|\bar{q}_2\gamma^\mu\gamma_5q_1|P\ra=i\sqrt{2}f_PP^\mu$
in Refs.~\cite{CJ1,CJ_PLB1}.}
Our model parameters and decay constants are summarized in
Table~\ref{t1} and compared with experimental data~\cite{PDG}
as well as lattice results~\cite{Ber}.
Note that in the numerical calculations we take $m_b=5.2$ GeV in all
formulas except in the Wilson coefficient $C^{\rm eff}_9$, where
$m_b=4.8$ GeV has been commonly used.

\begin{figure}
\centerline{\psfig{figure=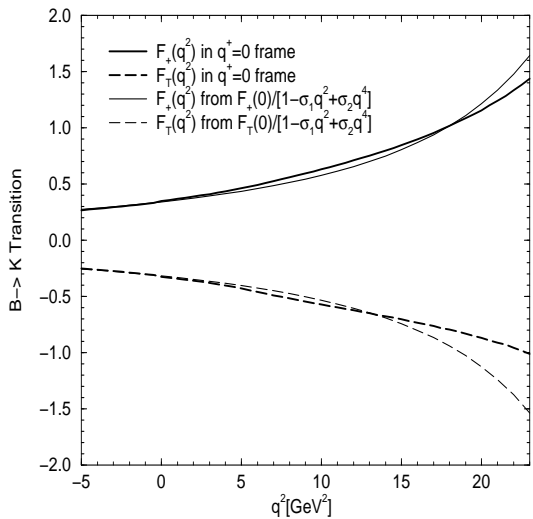,height=8cm,width=8cm}}
\caption{Analytic solutions of $F_+(q^2)$(thick solid line) and
$F_T(q^2)$(thick dashed line) compared with the results(thin lines)
obtained from the parametric formula given by Eq.~(\ref{Pole})
for $B\to K$ transition. \label{fig_FTP}}
\end{figure}

In Fig.~\ref{fig_FTP}, we show our analytic($q^+=0$ frame)
solutions for the weak form factors
$F_{+}(q^2)$(thick solid line) and $F_{T}(q^2)$(thick dashed line)
for $-5\; {\rm GeV}^2\leq q^2\leq (M_B-M_K)^2$. We also include
the results obtained from the parametric formula
given by Eq.~(\ref{Pole}) where the thin solid(dashed) line
represents $F_{+}(F_{T})$.
Our analytic solutions given by Eqs.~(\ref{FP}) and
(\ref{FT}) are well approximated by Eq.~(\ref{Pole}) up
to $q^2\lesssim 15$ GeV$^2$ but show some deviations near
zero recoil point.
We summarize in Table~\ref{t2} our numerical results for the weak form
factors $F_+(q^2)$ and $F_T(q^2)$ at $q^2=0$ and the parameters
$\sigma_i$ defined in Eq.~(\ref{Pole}) and compare with other theoretical
results~\cite{JW,Mel1,PB,AKS}. As one can see from Table~\ref{t2},
our results for the $F_+(q^2)$ and $F_T(q^2)$ in $q^2\to 0$ limit
are quite comparable with other theoretical results.
As other theoretical schemes predicted,
our results also show $F_{+}(0)(=0.348)\simeq -F_{T}(0)(=-0.324)$.

\begin{figure}
\centerline{\psfig{figure=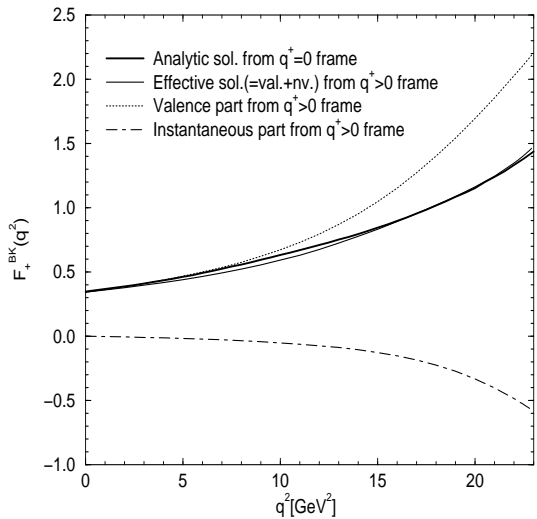,height=8cm,width=8cm}}
\caption{Effective solution of $F_+(q^2)$(thin solid line)
for $B\to K$ transition. The line code is in the figure.\label{fig_FPE}}
\end{figure}

For the analysis of heavy $\tau$ decay process, the weak form factor
$F_{-}(q^2)$(or equivalently $F_{0}(q^2)$) is necessary for the
calculations of the decay rate and the LPA and we obtain
it using our effective method~\cite{JC_E,CJK} in $q^+> 0$ frame as
described in Sec.~III(B).
In Fig.~\ref{fig_FPE}, we show our effective($q^+> 0$ frame)
solution of $F_{+}(q^2)$ (thin solid line) with a constant
$G_{BK}=3.9$ fixed by the normalization of $F_{+}(q^2)$ in
the $q^+=0$ frame (thick solid line) at $q^2=0$ limit.
As one can see in Fig.~\ref{fig_FPE}, our effective solution of
$F_{+}(q^2)$(thin solid line) is very close to the analytic
one(thick solid line) for the entire kinematic region. It justifies
the reliability of our constant approximation $G_{BK}$ of the kernel
${\cal K}$. For comparison, we also show the valence(dotted line) and
the instantaneous(dot-dashed line) contributions to $F_{+}(q^2)$
in the $q^+> 0$ frame.
Although the valence contribution dominates over the nonvalence
one for $q^2\lesssim 10$ GeV$^2$, the nonvalence (especially the
instantaneous) contribution is not negligible for
$q^2\gtrsim 10$ GeV$^2$.

\begin{figure}
\centerline{\psfig{figure=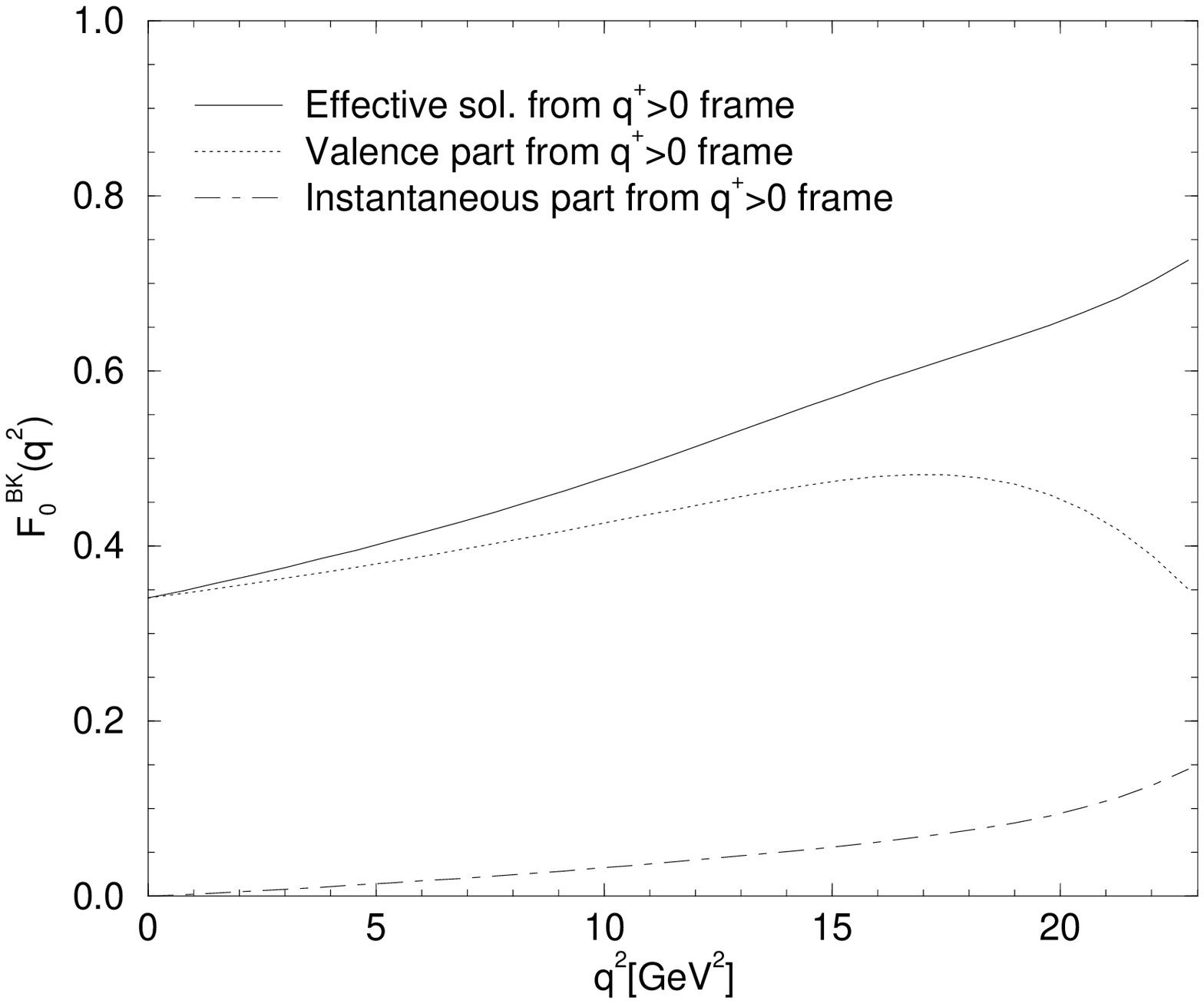,height=8cm,width=8cm}}
\vspace{-0.5cm}
\centerline{\psfig{figure=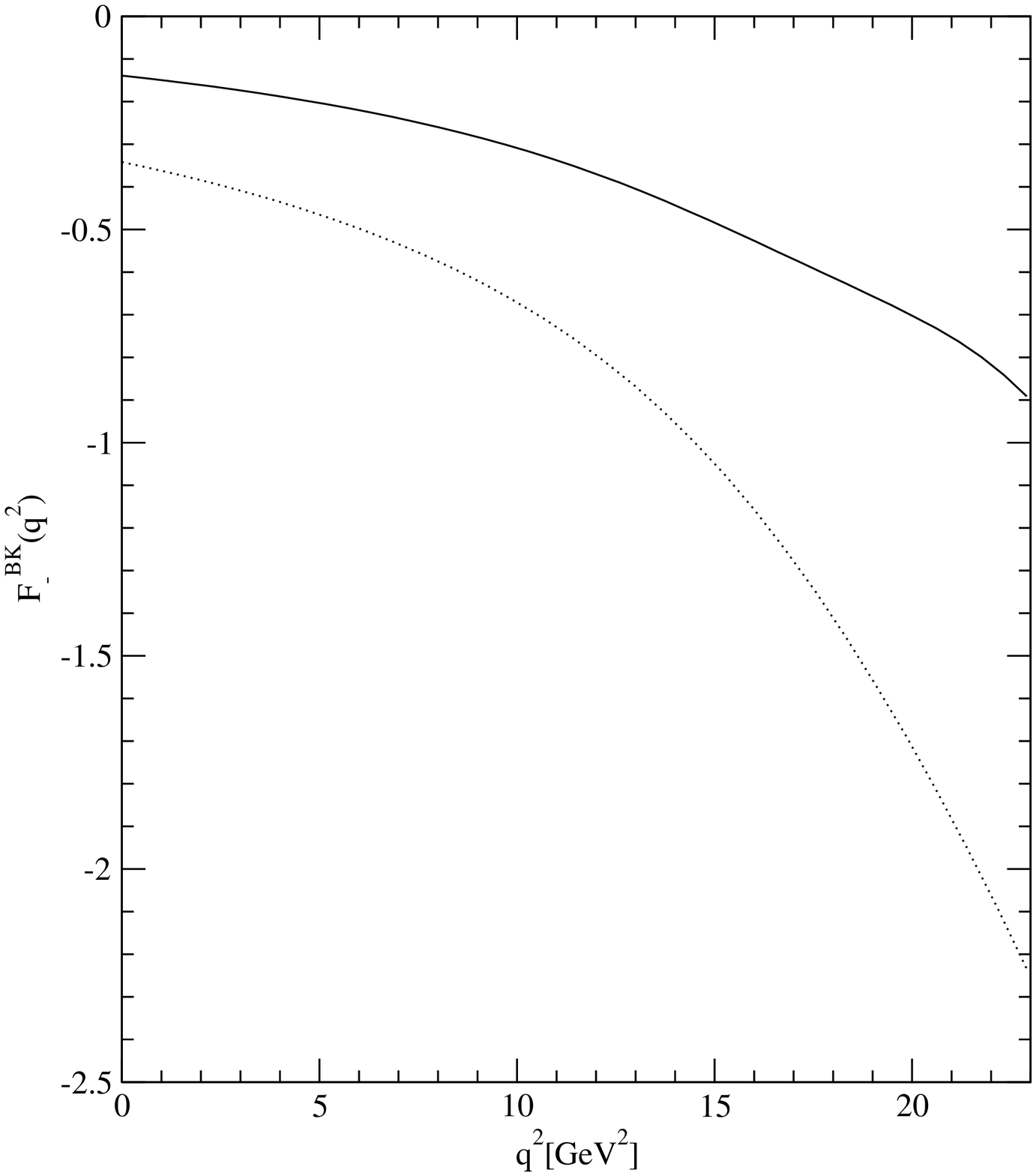,height=8cm,width=8cm}}
\caption{Effective solutions(solid line) of $F_0(q^2)$ and
$F_-(q^2)$ compared with the valence contributions(dotted line)
for $B\to K$ transition. \label{fig_F0E}}
\end{figure}

Using the same constant operator $G_{BK}=3.9$, we are now able to
calculate the scalar form factors $F_{0}(q^2)$
and $F_-(q^2)$ in $q^+>0$ frames
and the results are shown in Fig.~\ref{fig_F0E}(solid line).
As in the case of $F_{+}(q^2)$ in Fig.~\ref{fig_FPE}, we also
include the valence contributions(dotted line) to  both
$F_{0}(q^2)$ and $F_-(q^2)$ and the instantaneous
contribution(dot-dashed line) to $F_{0}(q^2)$.
It is very interesting to note especially from $F_-(q^2)$ that
the nonvalence contribution, i.e. the difference between solid
and dotted lines, is very substantial even at the maximum recoil
point($q^2=0$) and is growing as $q^2$ increases.
As a reference, our numerical results for $F_-$ obtained
from our effective(valence) solution at maximum- and zero-recoil
limits are $F_-(0)=-0.14(-0.34)$ and $F_-(q^2_{\rm max})=-0.9(-2.23)$,
respectively.
Our result for $F_-(q^2)$ presented in Fig.~\ref{fig_F0E} agrees very well
with the light cone QCD sum rule (LCSR) result for $F_-(q^2)$ by
Aliev et al.~\cite{AKOS}(See their Fig.1(b)).
Similarly, our effective solution for $F_0(q^2)$ is in a close agreement
with the LCSR results given by Ball~\cite{PB} and Ali et al.~\cite{ABHH}.
Our effective solution of $F_{0}(q^2)$ as well as the analytic solutions
of $F_{+}(q^2)$ and $F_{T}(q^2)$ shown in Fig.~\ref{fig_FTP}
will be used for the calculations of the branching ratios and the
longitudinal lepton polarization asymmetries.
We shall also discuss how we take the effect of the
vector meson dominance(VMD) into account at the end of this section.

\begin{figure}
\centerline{\psfig{figure=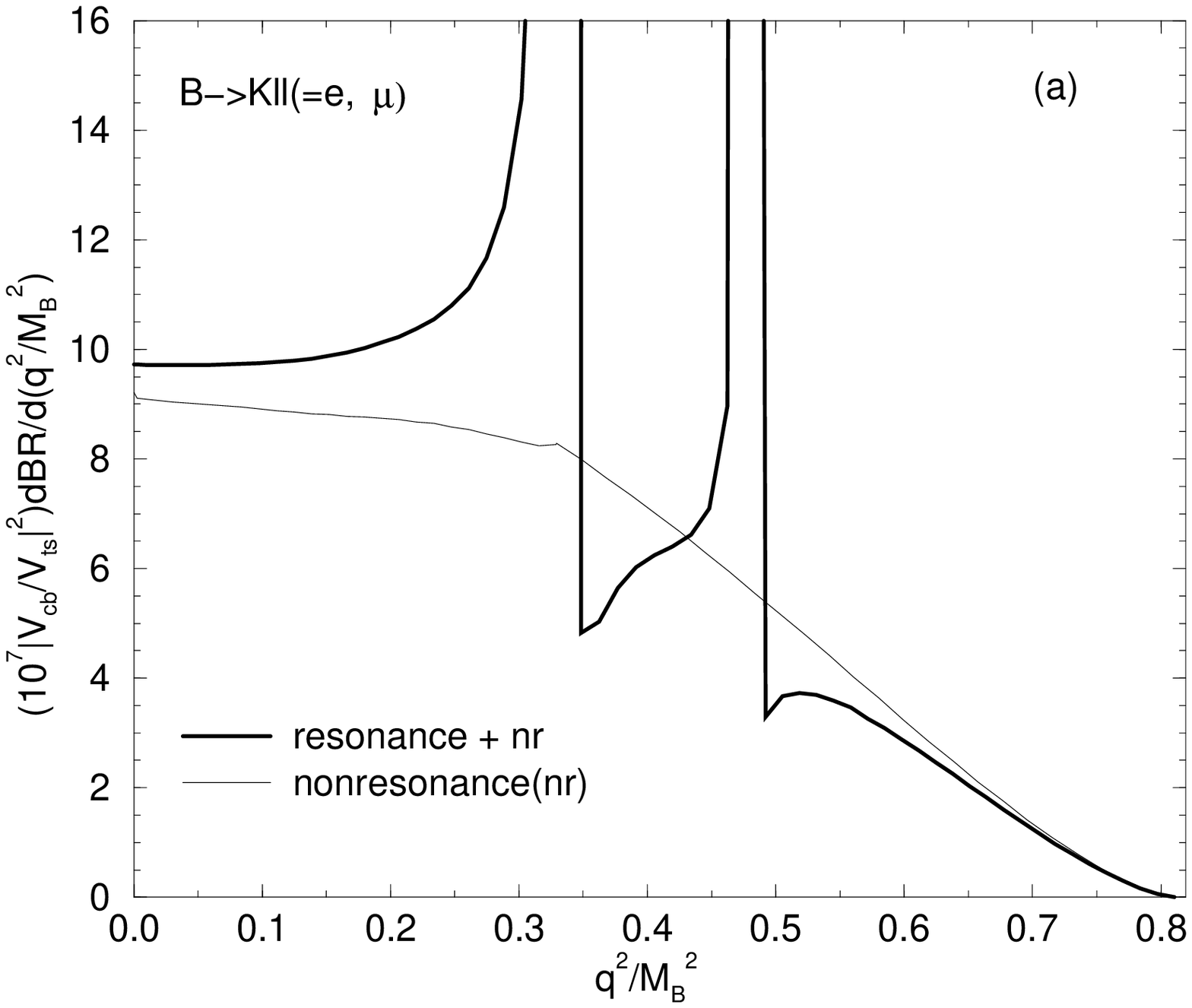,height=8cm,width=8cm}}
\centerline{\psfig{figure=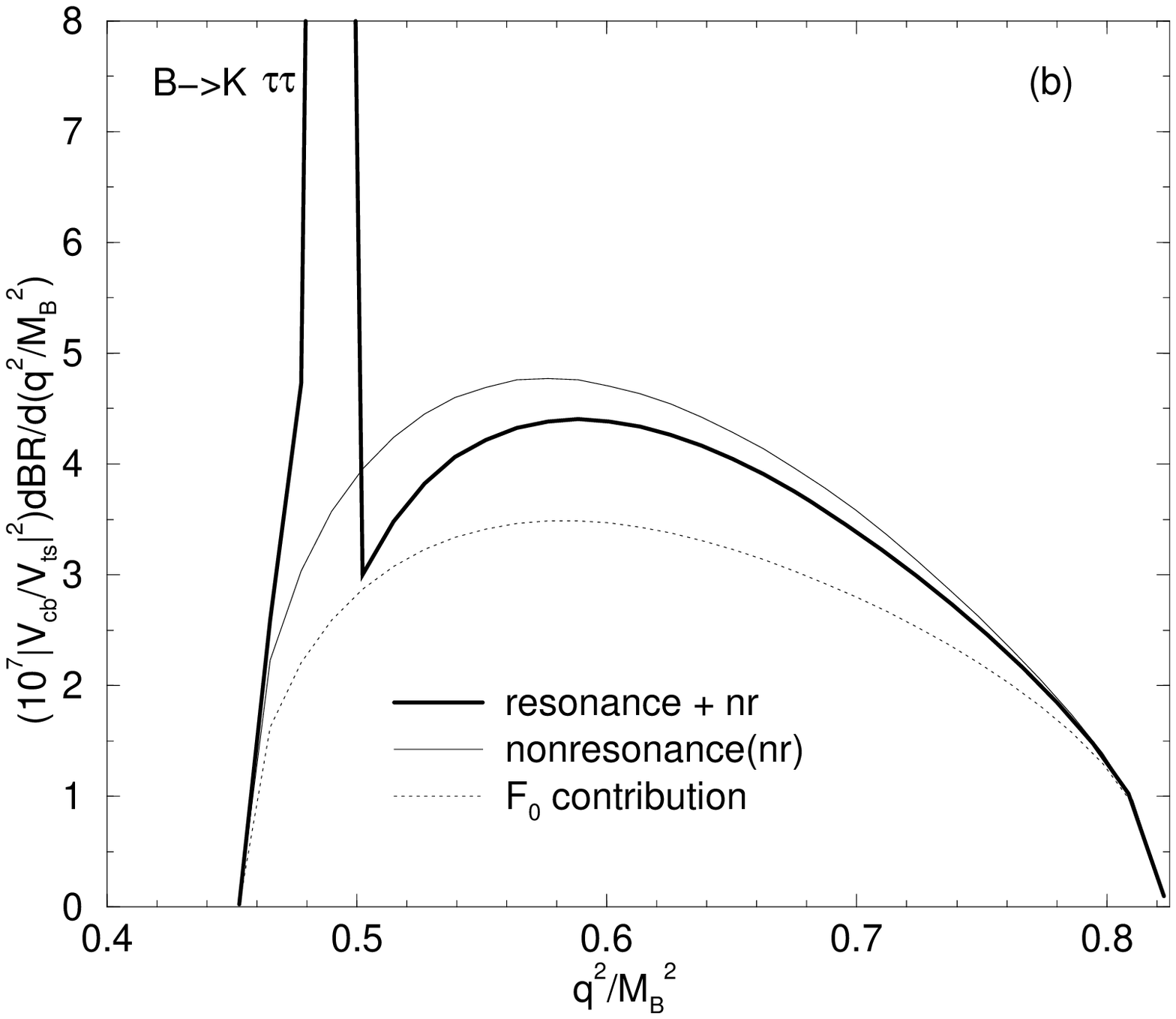,height=8cm,width=8cm}}
\caption{ The branching ratios for
$B\to K\ell^+\ell^-$($\ell=e,\mu$)(a) and $B\to K\tau^+\tau^-$(b)
transitions. The thick(thin) solid line represents the result
with(without) LD contribution to
$C^{\rm eff}_9$ in Eq.~(\protect\ref{C9}). The dotted line
in (b) represents the $F_0(q^2)$ contribution to the total branching
ratio of $\tau$ decay.\label{fig_BKrate}}
\end{figure}

We now show  our results for the differential branching ratios for
$B\to K\ell^+\ell^-(\ell=e,\mu)$ in Fig.~\ref{fig_BKrate}(a)
and $B\to K\tau^+\tau^-$ in Fig.~\ref{fig_BKrate}(b), respectively.
The thick(thin) solid line represents the result with(without)
the LD contribution($Y_{LD}(\hat{s})$) to $C^{\rm eff}_9$ given by
Eq.~(\ref{C9}).
In plotting Figs.~\ref{fig_BKrate}(a) and~(b), we set
$m_\ell=0$ and $m_{\tau}$=1.777 GeV, respectively.
As one can see the pole contributions clearly overwhelm the branching
ratio near $J/\psi(1S)$ and $\psi'(2S)$ peaks, however, suitable
$\ell^+\ell^-$ invariant mass cuts can separate the LD contribution
from the SD one away from these peaks.
This divides the spectrum into two distinct regions~\cite{Hew,AGM}:
(i) low-dilepton mass, $4m^2_\ell\leq q^2\leq M^2_{J/\psi}-\delta$,
and (ii) high-dilepton mass,
$M^2_{\psi'}+\delta\leq q^2\leq q^2_{\rm max}$, where $\delta$ is to
be matched to an experimental cut.
The branching ratios with[without] the pole(i.e. LD) contributions
for $B\to K\ell^+\ell^-$ are presented in Table~\ref{t3} for low(second
column), high(third column), and total(4th column) dilepton mass
regions of $q^2$.
Although the contribution of scalar form factor $F_0(q^2)$ to massless
lepton decay is negligible(zero for $m_\ell=0$), its
contribution to $\tau$-decay as shown in
Fig.~\ref{fig_BKrate}(b)(dotted line) is very substantial,
e.g. $\sim 75\%$ contribution to the total(nonresonant) decay rate
in our model calculation. Thus, the reliable calculation of
$F_{0}(q^2)$ is absolutely necessary and our effective method
of calculating the nonvalence diagram seems very useful.

It is worthwhile to compare our results for the branching ratios with
other light-front quark models\cite{GK,MN}. While the authors in
Ref.~\cite{GK} used the simple parametric formula, Eq.~(\ref{Pole}),
to obtain $F_{+}$ and $F_{T}$ and  the heavy quark symmetry(HQS)
to extract $F_{-}$, the authors in  Ref.~\cite{MN} used the dispersion
representation through the (Gaussian) wave functions of the initial and
final mesons and then analytically continue the form factors from
the spacelike region to the timelike region. The common aspect in these
models is to have
the same form factors $F_{+}$ and $F_{T}$, which are free from the zero-mode
contribution,  not in the timelike region but in the
spacelike region as far as the same model parameters are used.
Indeed our method of analytic continuation of the form factors $F_{+}$
and $F_{T}$ is equivalent to that of Ref.~\cite{MN}.
However, the difference is in the calculation of $F_{-}$, which
is not immune to the zero-mode contribution.
The zero-mode contribution must be properly taken into account
for the calculation of $F_{-}$.
Thus, it is not quite surprising to note that although our branching
ratio(see Fig.~\ref{fig_BKrate}(a)) for the
massless lepton $(\ell=e,\mu$) decay
is not much different from the results in Ref.~\cite{GK}(see
their Fig.~1(a)) and
Ref.~\cite{MN}(see their Fig.~3(a)), our branching ratio(see
Fig.~\ref{fig_BKrate}(b)) for
the $\tau$ decay is quite different from the results in
Ref.~\cite{GK}(see their Fig.~1(b)) and Ref.~\cite{MN}(see their Fig.~3(c)).

Our numerical results for the non-resonant branching ratios(assuming
$|V_{tb}|\simeq 1$) are $4.96\times 10^{-7}|V_{ts}/V_{cb}|^2$
for $B\to K\ell^+\ell^-$ ($\ell=e,\mu$) and
$1.27\times 10^{-7}|V_{ts}/V_{cb}|^2$ for
$B\to K\tau^+\tau^-$, respectively.
While the CLEO Collaboration~\cite{Bfactory}
reported the branching ratio ${\rm Br}(B\to K e^+e^-)<1.7\times 10^{-6}$,
the Belle Collaboration(K. Abe et al.)~\cite{Bfactory}
reported ${\rm Br}(B\to K e^+e^-)<1.2\times 10^{-6}$ and
${\rm Br}(B\to K\mu^+\mu^-)=(0.99^{+0.39+0.13}_{-0.32-0.15})\times 10^{-6}$,
respectively.
Our non-resonant results for the branching ratios of $B\to K\ell^+\ell^-$
is summarized in Table~\ref{t4} and compared with experimental data
as well as other theoretical predictions within the SM.


The exclusive $B\to K\tau^+\tau^-$ has been computed via
the heavy meson chiral perturbation theory by Du et al.~\cite{Du},
where the branching ratio of the exclusive decay was found to be about
$50-60\%$ of the inclusive one. Although calculations of
exclusive decay rates are inherently model dependent, chiral perturbation
theory is known to be reliable at energy scales smaller than the typical
scale of chiral symmetry breaking,
$\Lambda_{\rm CSB}\simeq 4\pi f_\pi/\sqrt{2}$. In $B\to K\tau^+\tau^-$, the
maximum energy of the $K$-meson in the $B$ rest frame is
$(M^2_B+M^2_K-4m^2_\tau)/2M_B\sim1.5$ GeV, which places most of the
available phase space around the scale
$\Lambda_{\rm CSB}$~\cite{Du,Hew}.
From the above argument and our exclusive $\tau$ branching fraction,
we can estimate the branching ratio of inclusive $B\to X_s\tau^+\tau^-$
as $(2.12-2.54)\times 10^{-7}|V_{ts}/V_{cb}|^2$ which is quite
comparable to the prediction given by Hewett~\cite{Hew} where
${\rm BR}(B\to X_s\tau^+\tau^-)=2.5\times 10^{-7}$ was obtained.

\begin{figure}
\centerline{\psfig{figure=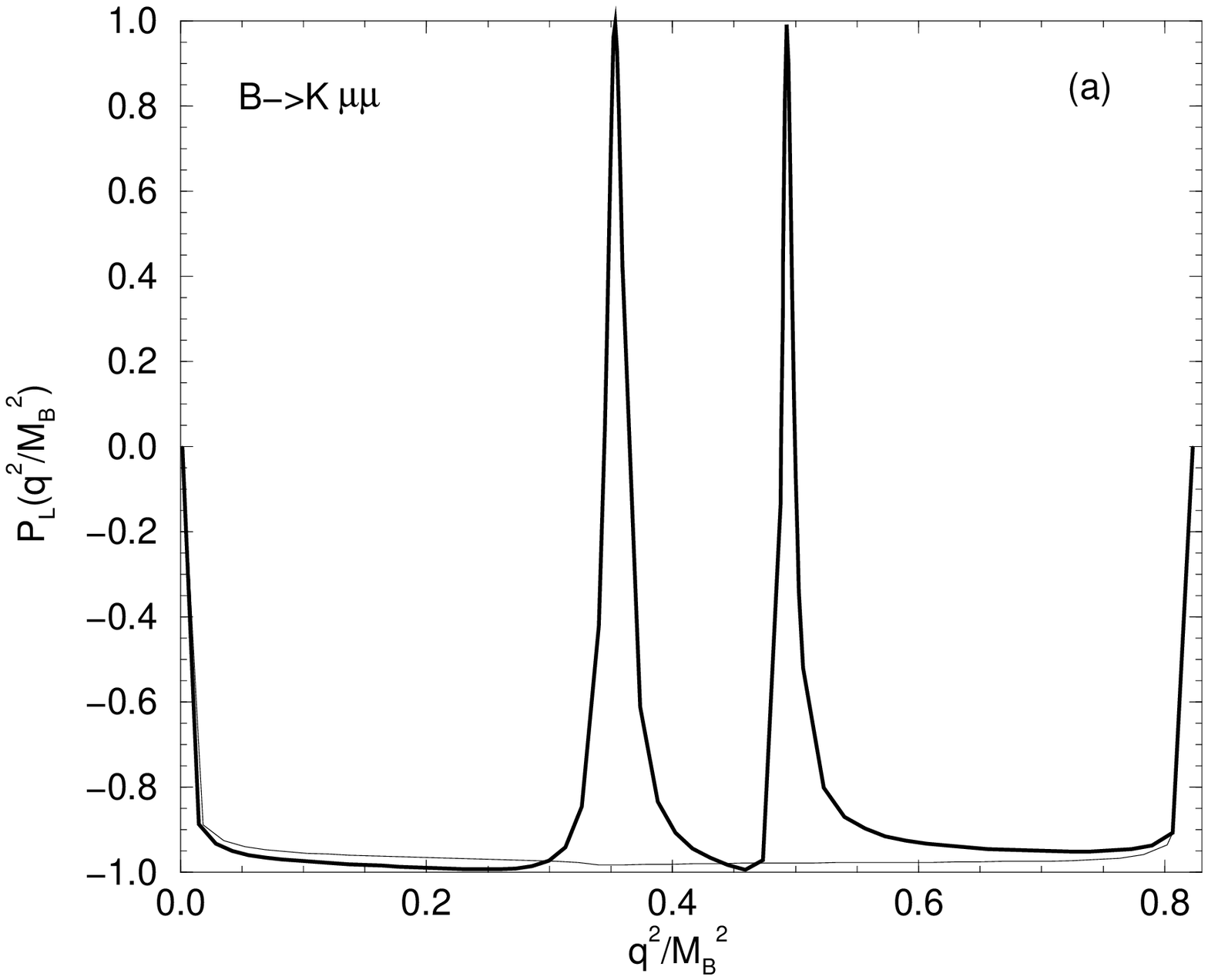,height=7.5cm,width=7.5cm}}
\centerline{\psfig{figure=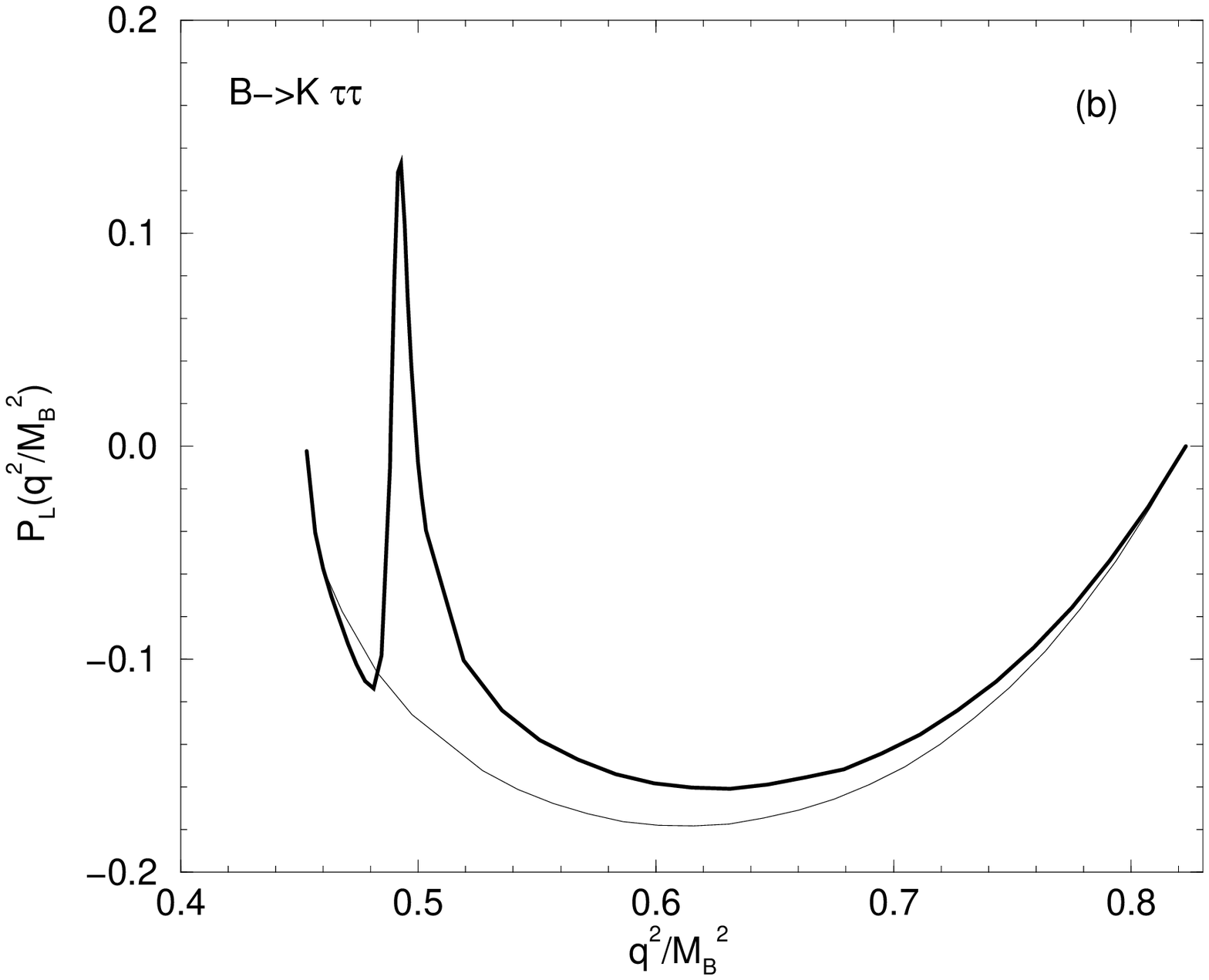,height=7.5cm,width=7.5cm}}
\caption{ The longitudinal lepton polarization asymmetries $P_L(\hat{s})$
for $B\to K\ell^+\ell^-$(a) and $B\to K\tau^+\tau^-$(b)
transitions. The same line code is used as in Fig.~\protect\ref{fig_BKrate}.
\label{fig_LPA}}
\end{figure}

In Figs.~\ref{fig_LPA}(a) and~(b), we show the
longitudinal lepton polarization asymmetries for $B\to K\mu^+\mu^-$
and $B\to K\tau^+\tau^-$ as a function of $\hat{s}$, respectively, and
with (thick solid line) and without (thin solid line) LD contributions.
For the $B\to K\mu^+\mu^-$ case,
we use the physical muon mass, $m_\mu$=105 MeV.
In both figures, the longitudinal lepton polarization asymmetries
become zero at the end point regions of $\hat{s}$.
Our numerical values of $P_L$ without LD contributions and away
from the end point regions are
$-0.97<P_L<-0.98$ in $0.3<\hat{s}<0.6$ region for $B\to K\mu^+\mu^-$ and
$-0.15<P_L<-0.18$ in $0.5<\hat{s}<0.7$ region for $B\to K\tau^+\tau^-$,
respectively.
In fact, the $P_L$ for the muon decay is insensitive to the form factors,
e.g. our $P_L\simeq -0.98$(away from the end points region) is well
approximated by~\cite{HQ}
\be\label{PLmu}
P_L\simeq 2\frac{C_{10}{\rm Re}C^{\rm eff}_9}{|C^{\rm eff}_9|^2
+ |C_{10}|^2}\simeq -1,
\ee
in the limit of $C_7\to 0$ from Eq.~(\ref{LPA_BK}). It also shows that
the $P_L$ for the $\mu$ dilepton channel
is insensitive to the little variation of $C_7$ as expected.
On the other hand, the LPA for the $\tau$ dilepton channel is sensitive
to the form factors. In other words, as in the case of branching ratios,
although our result of the LPA for the muon decay is not much different
from the results in Ref.~\cite{GK}(see their Fig.~2(a)) and
Ref~\cite{MN}(see their Fig.~5(a)),
the result for the tau decay is quite different
from the results in Ref.~\cite{GK}(see their Fig.~2(b)) and
Ref~\cite{MN}(see their Fig.~5(c)).

Comparing our results for the weak form factors with other phenomenological
models, one may find that there is in general a good agreement for
small and intermediate $q^2$ region. Nevertheless, there
are some differences for large $q^2$ region where
vector mesons are expected to dominate(VMD) especially for $F_+(q^2)$.
For example, both results of the LCSR
in~\cite{PB,BZ} and our LFQM analyses show that the direct
solution for $F_+(q^2)$ is well approximated by Eq.~(\ref{Pole}) up to
$q^2\lesssim 15$ GeV$^2$. However, the large momentum behavior of
$F_+(q^2)$(as well as $F_T(q^2)$) is somewhat different since our
model does not include the VMD effect.

Following the same method used in recent LCSR analysis~\cite{BZ},
we use the VMD formula(i.e. $B^*$-pole with
$M_{B^*}=5.325$ GeV) given by
\be\label{VMD}
F^{\rm VMD}_{+}(q^2)=\frac{c}{1-q^2/M^2_{B^*}}
\ee
at large $q^2$ region and match the parametric formula $F_+(q^2)$
in Eq.~(\ref{Pole}) by the following constraint~\cite{BZ}
\bea\label{inter}
F^{\rm VMD}_+(q^2)= F_+(q^2)
\;{\rm in}\; {\rm Eq.~(\ref{Pole})},
\nonumber\\
\frac{d}{dq^2}F^{\rm VMD}_+(q^2)=
\frac{d}{dq^2}F_+(q^2)\;{\rm in}\; {\rm Eq.~(\ref{Pole})},
\eea
to make both parametrizations smooth connection at a transition point
$q^2=q^2_0$, where $c$ is fixed at $q^2=q^2_0$ in Eq.~(\ref{inter}).
We should note that the $F_+(q^2)$ in Eq.~(\ref{Pole})
is almost equivalent to our LFQM prediction $F^{\rm LFQM}_+(q^2)$
up to $q^2\lesssim 15$ GeV$^2$ and the transition point $q^2_0$
is expected to be at $q^2\sim15$ GeV$^2$(see also Ref.~\cite{BZ})
in order to make interpolation between $F^{\rm LFQM}(q^2\leq q^2_0)$ and
$F^{\rm VMD}(q^2\geq q^2_0)$ more sense.
 \footnote{As discussed in~\protect\cite{Jaus96}, a naive extrapolation
of the VMD formula in Eq.~\protect(\ref{VMD}) to the point $q^2=0$ is
not consistent with the monopole formula
$F_+(q^2)=F_+(0)/(1-q^2/\Lambda^2_1)$ used in many theoretical
ansatz since the relevant parameters are in general different,
i.e. $F_+(0)\neq c$ and $\Lambda_1\neq M_{B^*}$.}
In our case for $B\to K$ transition, we obtain
$(c, q^2_0)$=(0.388, 14.38 GeV$^2$) for $F^{\rm BK}_+(q^2)$.
For the tensor form factor,
we get $(c, q^2_0)$=($-0.358$, 14.23 GeV$^2$) for $F_T(q^2)$.

It is necessary to discuss the exclusive
$B\to\pi\ell\nu_\ell$ process in that the constant $c$ has a direct
physical implication for $B\to\pi\ell\nu_\ell$ process, i.e. it
is related to the physical couplings as~\cite{BZ,BBKR,KRWY}
\be\label{c_phy}
c=\frac{f_{B^*}g_{B^*B\pi}}{2M_{B^*}}
\ee
where $f_{B^*}$ is the decay constant of the $B^*$ meson defined by
$\la 0|\bar{b}\gamma^\mu u|B^*\ra=M_{B^*}f_{B^*}\epsilon^\mu$
and $g_{B^*B\pi}$ is the (axial-current) coupling defined by
$\la B^0(P)\pi^+(q)|B^{*+}(P+q)\ra=g_{BB^*\pi}(q\cdot\epsilon)$
and can be extracted from soft pion $q^2\to 0$ limit in the heavy
meson chiral perturbation theory~\cite{WBW,WBW2}.
In the limit where the heavy quark mass $m_Q(Q=c,b)$ goes to infinity
there are flavor-independent relations between coupling constants
\be\label{HQS}
g=\frac{f_\pi}{2M_D}g_{D^*D\pi}
=\frac{f_\pi}{2M_B}g_{B^*B\pi},
\ee
where $f_\pi=131$ MeV and the coupling constant $g$ appears in the
interaction Lagrangian of the effective meson field
theory~\cite{Cas,WBW,WBW2}.

In our numerical calculation of $c$ for the exclusive $B\to\pi e\nu_e$
process, we obtain ($c,q^2_0$)=(0.312,15.12 GeV$^2$) from Eq.~(\ref{inter})
and $(\sigma_1,\sigma_2)=(4.75\times 10^{-2},5.50\times 10^{-4})$ in
Eq.~(\ref{Pole}), which was obtained in
our previous analysis~\cite{thesis}.
Since we also obtained the $B^*$ meson decay constant
as $f_{B^*}=185.8$ MeV~\cite{thesis}, we can now extract
the coupling constant of the $B^*$ to $B\pi$-pair and the result
is $g_{B^*B\pi^+}=17.88$ and $g$=0.23 while the recent
fit~\cite{IWS} to the experimental data gives two possible solutions,
$g=0.27^{+0.04+0.05}_{-0.02-0.02}$ or
$g=0.76^{+0.03+0.2}_{-0.03-0.1}$. We acknowledge the
remark in~\cite{IWS} that for the $B\to\pi\ell\nu_\ell$ form factors
with $E_\pi<2m_\pi$, analytic bounds combined with chiral perturbation
theory give $gf_B\lesssim 50$ MeV~\cite{Boyd}. That means while
the solution $g=0.27$ gives $f_B\lesssim 190$ MeV, $g=0.76$ gives
$f_B\lesssim 66$ MeV, which is roughly a factor of three smaller
than lattice QCD result~\cite{Ber},{\it i.e.}
$f^{\rm Lat.}_B= 200\pm 30$ MeV. Note that our LFQM prediction is
given by $f^{\rm LFQM}_B=171.4$ MeV.
As a reference, other theoretical calculations for $g$ are
$0.2-0.4$ for the QCD sum rules, $1/3-0.6$ for the quark
models\footnote{Using similar LFQM to ours, Jaus~\cite{Jaus96}
obtained $g=0.56$ from the direct calculation of the hadronic
matrix element in the soft pion limit and argued that
the calculated $\rho - \pi - \pi$ and $K^* - K - \pi$ coupling constants 
within the same model are in fair agreement with data. The reason for the 
discrepancy of $g$ value is not yet understood. 
However, the computed decay constants 
$f_B$ and $f_{B^*}$ are in good agreement between Ref.\cite{Jaus96} and 
ours.} and 0.42(4)(8) for the lattice calculation(see Ref.~\cite{BY} for the 
survey of $g$ values obtained from different models).

\begin{figure}
\centerline{\psfig{figure=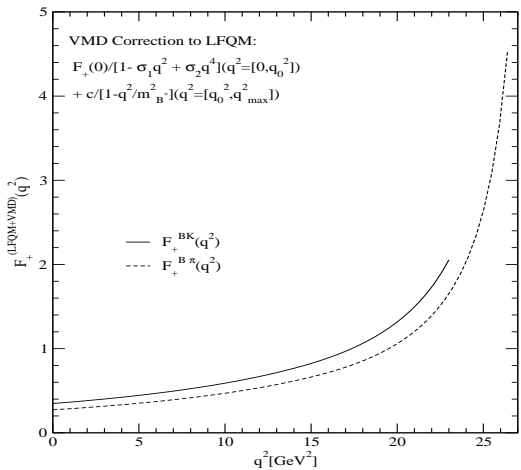,height=8cm,width=8cm}}
\caption{ VMD corrections to the LFQM predctions for
$F^{ BK}_+(q^2)$(solid line) and $F^{ B\pi}_+(q^2)$(dashed line),
i.e. $F_+(q^2)=F^{\rm LFQM}_+(q^2\leq q^2_0)
+F^{\rm VMD}_+(q^2\geq q^2_0)$.
\label{fig_9}}
\end{figure}

In Fig.~\ref{fig_9}, we show the VMD corrections to both
$F^{BK}_+(q^2)$(solid line) and $F^{B\pi}_+(q^2)$(dashed line),
i.e. $F_+(q^2)=F^{\rm LFQM}_+(q^2\leq q^2_0)
+F^{\rm VMD}_+(q^2\geq q^2_0)$.
Comparing Fig.~\ref{fig_FTP}[Fig.~3 in~\cite{CJ_PLB1}]
and Fig.~\ref{fig_9}, we find the enhancement of
$F^{BK}_+(q^2)[F^{B\pi}_+(q^2)]$ at $q^2=q^2_{\rm max}$
by around $40[70]\%$. Our result for $F^{B\pi}_+(q^2)$
including the VMD correction are quite comparable with that
obtained from QCD sum rules in Ref.~\cite{BZ} where the authors
used the same method to enhance $F^{B\pi}_+(q^2)$.
Our result for $F^{BK}_+(q^2)$ in Fig.~\ref{fig_9} is also
comparable with those of Refs.~\cite{PB,ABHH}.
However, the branching ratio for $B\to K\ell^+\ell^-$($\ell=e,\mu$)
increases less than $2\%$ by including the VMD effect.
It is not surprising to note that the large enhancement of the
weak form factors near the zero-recoil($q^2=q^2_{\rm max}$) region
does not affect the differential decay rate very much, since the
phase space of the large $q^2$ region is highly suppressed in
Eq.~(\ref{DDR}).

\section{Summary and Conclusion}
In this work, we investigated the rare exclusive semilpetonic
$B\to K\ell^+\ell^-$ ($\ell=e,\mu$ and $\tau$) decays within the
SM, using our LFQM which
has been tested extensively in spacelike processes~\cite{CJ1,CJK}
as well as in the timelike exclusive semileptonice decays
of pseudoscalar mesons~\cite{CJ_PLB1,JC_E}.
The form factors $F_{+}(q^2)$ and $F_T(q^2)$ are obtained
in the $q^{+}=0$ frame ($q^2<0$) and then analytically
continued to the timelike region by changing $q_{\perp}$ to
$iq_{\perp}$ in the form factors.
The form factor $F_-(q^2)$ is obtained from our effective
treatment of the nonvalence contribution in addition to the
valence one in $q^+>0$ frames ($q^2>0$) based on the B-S
formalism.  The covariance (i.e. frame-independence) of our model
has been checked by comparison of $F_+(q^2)$ obtained from both
$q^+=0$ and $q^+>0$ frames. Our numerical results for the form factors
are comparable with other theoretical calculations as shown in
Table~\ref{t2}.
Using the solutions of $F_+$ and $F_T$ obtained from $q^+=0$ frame
and $F_-$ obtained from $q^+>0$ frame, we calculate the branching
ratios and the longitudinal lepton polarization asymmetries
for $B\to K\ell^+\ell^-$ including both
short- and long-distance contributions from QCD Wilson coefficients.
Our numerical results for the non-resonant branching ratios are
in the order of $10^{-7}$, which are consistent with many other
theoretical predictions as shown in Table~\ref{t4}.
Of particular interest, we were able to estimate the inclusive
branching ratio for $B\to X_s\tau^+\tau^-$ as
${\rm BR}(B\to X_s\tau^+\tau^-)\sim(2.12-2.54)\times 10^{-7}|V_{ts}/V_{cb}|^2$
with the help of chiral perturbation theory~\cite{Du}.
For the LPA as a parity-violating observable, we find
that the LPA for the $\tau$ channel is sensitive to the form factors
while the LPA for the $\mu$ channel is insensitve to the
model for the hadronic form factors.
Thus, the experimental data of the LPA for $\tau$ decay would provide a
useful guidance for the model building of hadrons and make a definitive test
on existing models.
\acknowledgements
The work of HMC and LSK was supported in part by the NSF
grant PHY-00070888 and that of CRJ by the US DOE under grant
No. DE-FG02-96ER40947. The North Carolina Supercomputing Center and
the National Energy Research Scientific Computer Center are also
acknowledged for the grant of Cray time.

\setcounter{equation}{0}
\setcounter{equation}{0}
\renewcommand{\theequation}{\mbox{A\arabic{equation}}}
\begin{center}
{\bf APPENDIX A: FUNCTIONS $Y_{\rm SD}(\hat{s})$,
$Y_{\rm LD}(\hat{s})$, AND $\omega(\hat{s})$
in Eq.~(\ref{C9})}
\end{center}
The function $Y_{\rm SD}(\hat{s})$ in Eq.~(\ref{C9}) is given by
\bea\label{YSD}
Y_{\rm SD}(\hat{s})&=& h(\hat{m}_c,\hat{s})
(3C_1+C_2+C^{(0)})
\nonumber\\
&&-\frac{1}{2}h(1,\hat{s})(4C_3 + 4C_4+3C_5+C_6)
\nonumber\\
&&-\frac{1}{2}h(0,\hat{s})(C_3+3C_4)
+\frac{2}{9}C^{(0)}
\nonumber\\
&&-\frac{V^*_{us}V_{ub}}{V^*_{ts}V_{tb}}(3C_1+C_2)
[h(0,\hat{s})-h(\hat{m}_c,\hat{s})],
\eea
where $C^{(0)}\equiv 3C_3+C_4+3C_5+C_6$.
The function $h$($\hat{m}_q$=$m_q/m_b$,$\hat{s}$) in
Eq.~(\ref{YSD}) arises from the one loop contributions of the
four quark operators $O_1-O_6$
and $h(\hat{m}_c,\hat{s}),h(1,\hat{s})$, and
$h(0,\hat{s})$ represent $c$ quark, $b$ quark, and $u,d,s$ quark
loop contributions, respectively. The explicit form of
$h(\hat{m}_q,\hat{s})$ is given by
\bea\label{gms}
h(\hat{m}_q,\hat{s})&=&
-\frac{8}{9}{\rm ln}\biggl(\frac{m_b}{\mu}\biggr)
-\frac{8}{9}{\rm ln}\hat{m}_q
+ \frac{8}{27} + \frac{4}{9}y_q
\nonumber\\
&&-\frac{2}{9}(2+y_q)\sqrt{|1-y_q|}
\nonumber\\
&&\times\biggl\{
\Theta(1-y_q)
\biggl[
{\rm ln}\frac{ 1+ \sqrt{1-y_q} }{ 1-\sqrt{1-y_q} } - i\pi
\biggr]
\nonumber\\
&&\hspace{0.5cm} + \Theta(y_q - 1)
2{\rm arctan}\frac{1}{ \sqrt{y_q - 1} }
\biggr\},
\eea
where $y_q=4\hat{m}^2_q/\hat{s}$ and
\be\label{g0s}
h(0,\hat{s})=\frac{8}{27}
-\frac{8}{9}{\rm ln}\biggl(\frac{m_b}{\mu}\biggr)
-\frac{4}{9}{\rm ln}\hat{s}
+\frac{4}{9}i\pi.
\ee
The function $Y_{\rm LD}(\hat{s})$ in Eq.~(\ref{C9}) is given by
\bea\label{YLD}
Y_{\rm LD}(\hat{s})&=&\frac{3\kappa}{\alpha^2}
\biggl[-\frac{V^*_{cs}V_{cb}}{V^*_{ts}V_{tb}}
(3C_1+C_2+C^{(0)})
-\frac{V^*_{us}V_{ub}}{V^*_{ts}V_{tb}}C^{(0)}\biggr]
\nonumber\\
&&\times\sum_{V_i=J/\psi,\psi',\ldots}
\frac{\pi\Gamma(V_i\to\ell^+\ell^-)M_{V_i}}
{M^2_{V_i}-\hat{s}m^2_b-iM_{V_i}\Gamma_{V_i}},
\eea
where $\Gamma(V_i\to\ell^+\ell^-)$, $\Gamma_{V_i}$ and $M_{V_i}$ are the
leptonic decay rate, width and mass of the
$i$th $1^{--}$ $c\bar{c}$ resonance, respectively. In our numerical
calculations, we use $\Gamma(J/\psi\to\ell^+\ell^-)=5.26\times 10^{-6}$
GeV, $M_{J/\psi}=3.1$ GeV, $\Gamma_{J/\psi}=87\times 10^{-6}$ GeV
for $J/\psi(1S)$ and $\Gamma(\psi'\to\ell^+\ell^-)=2.12\times 10^{-6}$
GeV, $M_{\psi'}=3.69$ GeV, $\Gamma_{\psi'}=277\times 10^{-6}$ GeV
for $\psi'(2S)$~\cite{PDG}.
The fudge factor $\kappa$ is introduced in Eq.~(\ref{YLD}) to account
for inadequacies of the naive factorization framework
(see~\cite{MS} for more details.)
We adopt $\kappa$=2.3~\cite{LW} to reproduce the rate of decay
chain $B\to X_s J/\psi\to X_s\ell^+\ell^-$.

In the SD contribution of $b\to s\ell^+\ell^-$, the $u$-quark
loop contribution is neglected due to the smallness of the
contribution $V^*_{us}V_{ub}/V^*_{ts}V_{tb}\simeq O(\lambda^2)$
($\lambda\simeq 0.22$ is Wolfenstein parameter) compared
with $V^*_{cs}V_{cb}\simeq -V^*_{ts}V_{tb}$.
The term $(V^*_{us}V_{ub}/V^*_{ts}V_{tb})C^{(0)}$ in
LD contribution is also neglected for $b\to s\ell^+\ell^-$.

The function $\omega(\hat{s})$ in Eq.~(\ref{C9}) represents
the $O(\alpha_s)$ correction from the one-gluon exchange in
the matrix element of $O_9$~\cite{JK}:
\bea\label{WS}
\omega(\hat{s})&=&-\frac{2}{9}\pi^2 -\frac{4}{3}Li_2(\hat{s})
-\frac{2}{3}{\rm ln}\hat{s}{\rm ln}(1-\hat{s})
\nonumber\\
&&-\frac{ 5 + 4\hat{s} }{ 3(1+2\hat{s}) }{\rm ln}(1-\hat{s})
-\frac{ 2\hat{s}(1+\hat{s})(1-2\hat{s}) } { 3(1-\hat{s})^2(1+2\hat{s}) }
{\rm ln}\hat{s}
\nonumber\\
&&+\frac{ 5 + 9\hat{s} - 6\hat{s}^2 }{ 6(1-\hat{s})(1+2\hat{s}) },
\eea
where $Li_2(x)=-\int^{1}_0 dt\;{\rm ln}(1-xt)/t$.
\setcounter{equation}{0}
\renewcommand{\theequation}{\mbox{B\arabic{equation}}}
\begin{center}
{\bf APPENDIX B: DERIVATION OF THE DECAY RATE FOR $B\to K\ell^+\ell^-$ }
\end{center}

In this appendix, we show the derivation of
the decay rate for $B\to K\ell^+\ell^-$.
For simplicity, we shall omit the factor $V^*_{ts}V_{tb}$
in the following derivation.

The transition amplitude for $B\to K\ell^+\ell^-$ is given by
\bea\label{tamp}
{\cal M} &=& \la K\ell^+\ell^-|{\cal H}|B\ra \nonumber\\
&=&\frac{4G_F}{\sqrt{2}}\frac{\alpha}{4\pi}
\biggl\{\biggl[C^{\rm eff}_9 J_\mu - \frac{2m_b}{q^2}C_7 J^T_\mu\biggr]
\bar{\ell}\gamma^\mu\ell \nonumber\\
&&\hspace{1.5cm}
+ C_{10}J_\mu \bar{\ell}\gamma^\mu\gamma_5\ell \biggr\}.
\eea
For all possible spin configurations, we make the replacement
\be\label{atamp}
|{\cal M}|^2\to
\overline{|{\cal M}|^2}\equiv\frac{1}{(2S_B+1)(2S_K+1)}
\sum_{\rm all\;spin\;states}|{\cal M}|^2,
\ee
where $S_B(S_K)$ is the spin of $B(K)$ meson and we
sum over the spins of the lepton pair.
After summing over all spin states for the lepton pair,
we obtain
\bea\label{ass}
\overline{|{\cal M}|^2}
&=& \frac{G^2_F}{2\pi^2}\alpha^2
\biggl[ [2(P\cdot p_\ell)(P\cdot p_{\bar{\ell}})
-\frac{P^2q^2}{2}]F_{T+}
\nonumber\\
&&\hspace{1.3cm}
+2\frac{\hat{m}_\ell}{\hat{s}}{\cal F}_{0+}
\biggr],
\eea
where $F_{T+}$ is given by Eq.~(\ref{DDR2}) and
\be\label{F0+}
{\cal F}_{0+}=
|C_{10}|^2\biggl( [q^2P^2-(P\cdot q)^2]|F_+|^2
+ (P\cdot q)^2|F_0|^2 \biggr).
\ee
Here, we use $m_b\simeq M_B$ in the derivation of Eq.~(\ref{ass}).

In the $B-$meson rest frame, Eq.~(\ref{ass}) can be rewritten as
\bea\label{ass2}
\overline{|{\cal M}|^2}
&=& \frac{M^2_B G^2_F}{\pi^2}\alpha^2
\biggl[
[ |\vec{P}_K|^2 - (E_\ell - E_{\bar{\ell}})^2 ] F_{T+}
\nonumber\\
&&\hspace{1.8cm}
+\frac{\hat{m}_\ell}{\hat{s}}M^2_B F_{0+}
\biggr],
\eea
where $|\vec{P}_K|^2=M^2_B \hat{\phi}/4$.

The differential decay rate for $B\to K\ell^+\ell^-$ is given
by
\bea\label{DrateBK}
d\Gamma &=&
\frac{\overline{|{\cal M}|^2}}{2M_B}
\biggl( \frac{d^3\vec{P}_K}{(2\pi)^3 2E_K}\biggr)
\biggl( \frac{d^3\vec{P}_\ell}{(2\pi)^3 2E_\ell}\biggr)
\biggl( \frac{d^3\vec{P}_{\bar{\ell}} }{(2\pi)^3 2E_{\bar{\ell}}} \biggr)
\nonumber\\
&&\times
(2\pi)^4\delta^4(P_B-P_K-P_\ell-P_{\bar{\ell}}).
\eea
After doing the $\vec{P}_{\bar{\ell}}$ integration, one obtains
\bea\label{DrateBK2}
d\Gamma
&=& \frac{M_B G^2_F}{64\pi^5}\alpha^2
\biggl[ [|\vec{P}_K|^2-(2E_\ell + E_K-M_B)^2]F_{T+}
\nonumber\\
&&\hspace{1.8cm}
+\frac{\hat{m}_\ell}{\hat{s}}M^2_B F_{0+}
\biggl]dE_K dE_\ell.
\eea
The lepton energy $E_\ell$ in Eq.~(\ref{DrateBK2})
satisfies the following upper($E^+_\ell$) and lower($E^-_\ell$) bounds
\be\label{EPM}
E^{\pm}_\ell=
\frac{(M_B-E_K)\pm|\vec{P}_K|\sqrt{1-4(\hat{m}_\ell/\hat{s})}}
{2}.
\ee
Finally, the integration of Eq.~(\ref{DrateBK2}) over $E_\ell$ with
$dE_K=(M_B/2)d\hat{s}$ gives Eq.~(\ref{DDR}).

\setcounter{equation}{0}
\renewcommand{\theequation}{\mbox{C\arabic{equation}}}
\begin{center}
{\bf APPENDIX C: ANALYTIC FORM OF THE WEAK FORM FACTORS IN TIMELIKE REGION }
\end{center}
In this appendix, we show the generic form of
our analytic solutions for the weak
form factors $F_+(q^2)$[Eq.~(\ref{FP})] and
$F_T(q^2)$[Eq.~(\ref{FT})] in timelike region.

In our numerical analysis, we use change of variables as
\bea\label{Cv}
\vec{k}_\perp&=&\vec{\ell}_\perp
+\frac{x\beta^2_1}{\beta^2_1+\beta^2_2}\vec{q}_\perp,
\nonumber\\
\vec{k'}_\perp&=&\vec{\ell}_\perp
-\frac{x\beta^2_2}{\beta^2_1+\beta^2_2}\vec{q}_\perp.
\eea
Since the form factors in Eqs.~(\ref{FP}) and~(\ref{FT}) involve
the terms proportional to $(\vec{\ell}_\perp\cdot\vec{q}_\perp)^{\rm odd}$,
which are related to the imaginary parts of the form factors by
changing $\vec{q}_\perp$ to $i\vec{q}_\perp$, we separate the terms
with even powers of ($\vec{\ell}^2_\perp,\vec{q}^2_\perp$) from
those with $(\vec{\ell}_\perp\cdot\vec{q}_\perp)^{\rm odd}$
in the form factors.
One useful identity in this separation procedure is
\bea\label{pq_L}
\sqrt{2}\sqrt{a + b(\vec{p}_\perp\cdot\vec{q}_\perp)}
&=& \sqrt{a + \sqrt{a^2-b^2(\vec{p}_\perp\cdot\vec{q}_\perp)^2}}\nonumber\\
&&+ \frac{b(\vec{p}_\perp\cdot\vec{q}_\perp)}
{\sqrt{a + \sqrt{a^2-b^2(\vec{p}_\perp\cdot\vec{q}_\perp)^2}} }.
\eea
By changing $\vec{p}_\perp\cdot\vec{q}_\perp\to
i\vec{p}_\perp\cdot\vec{q}_\perp
=i|\vec{\ell}_\perp|\sqrt{q^2}\cos\theta\equiv i\delta_l$ where
$q^2>0$,
we separate the `Real'-parts from `Imaginary'-parts in
Eqs.~(\ref{FP}) and~(\ref{FT}) as follows
\bea\label{anale}
\frac{\beta^2_1\vec{k'}^2_2+\beta^2_2\vec{k}^2_1}{2\beta^2_1\beta^2_2}
&\equiv&
\bar{\ell}_R(\vec{\ell}^2_\perp,q^2)
+ i\delta_l\bar{\ell}_I(\vec{\ell}^2_\perp,q^2),
\eea
from the exponent of $\phi_2\phi_1$, and
\bea\label{analj}
\sqrt{\frac{\partial k'_z}{\partial x}}
\sqrt{\frac{\partial k_z}{\partial x}}&\equiv&
{\cal J}_R(\vec{\ell}^2_\perp,q^2)
+ i\delta_l{\cal J}_I(\vec{\ell}^2_\perp,q^2),
\eea
from the Jacobi factor. The separations of Eqs.~(\ref{anale})
and~(\ref{analj}) are common for both $F_+(q^2)$ and $F_T(q^2)$.
The main difference between the two form factors comes from different
vertex structure and we denote generically as
\bea\label{analm}
&&\sum_{{\lambda}'s}{\cal R}^{00^\dagger}_{\lambda_2\bar{\lambda}}
\frac{\bar{u}_{\lambda_2}(p_2)}{\sqrt{p^+_2}}\Gamma^+
\frac{u_{\lambda_1}(p_1)}{\sqrt{p^+_1}}
{\cal R}^{00}_{\lambda_1\bar{\lambda}}\nonumber\\
&&\;= {\cal M}_R(\vec{\ell}^2_\perp,q^2)
+ i\delta_l{\cal M}_I(\vec{\ell}^2_\perp,q^2).
\eea
Combining Eqs.~(\ref{anale}-\ref{analm}), we separate the
`Real' and `Imaginary' parts of the weak form factors:
\bea\label{analf}
F(q^2)&=&\frac{1}{(\pi\beta_1\beta_2)^{3/2}}
\int^1_0 dx\int d^2\vec{\ell}_\perp
\exp(-\bar{\ell}_R)
\nonumber\\
&\times&
\biggl[
[{\cal J}_R{\cal M}_R-\delta^2_l{\cal J}_I{\cal M}_I]
[\cos(\delta_l\bar{\ell}_I) - i\sin(\delta_l\bar{\ell}_I)]
\nonumber\\
&&\;\;
+\delta_l[{\cal J}_R{\cal M}_I + {\cal J}_I{\cal M}_R]
[\sin(\delta_l\bar{\ell}_I) + i\cos(\delta_l\bar{\ell}_I)]
\biggr],\nonumber\\
&\equiv&F_R(q^2) + i F_{Im}(q^2).
\eea
We do not list here the detailed functional forms of other terms.
However, since only the term $\delta_l$ is of odd power in $\vec{\ell}_\perp$
and $\vec{q}_\perp$, one can easily check the imaginary term
of the form factor $F_{Im}(q^2)$ vanishes after $\ell_\perp$ integration
due to the fact that
$\int d^2\vec{\ell}_\perp\ell^{\rm odd}_\perp
\exp(-\ell^{\rm even}_\perp)=0$.
In fact, we also found that the term $\delta_l\bar{\ell}_I$ is small enough
to make $\cos(\delta_l\bar{\ell}_I)\simeq 1$ and
$\sin(\delta_l\bar{\ell}_I)\simeq\delta_l\bar{\ell}_I$ with very high
accuracy.

\begin{table}
\caption{Model Parameters ($m_q,\beta$)
and the decay constants
defined by $\la 0|\bar{q}_2\gamma^\mu\gamma_5q_1|P\ra=if_PP^\mu$
for $\pi$, $K$ and $B$ mesons used in our analysis. We also
compare our decay constants with the data~\protect\cite{PDG}
and the lattice result~\protect\cite{Ber}. }
\begin{tabular}{c|c|c|c|c}
Meson($q\bar{Q}$) & $m_Q$[GeV] & $\beta_{q\bar{Q}}$[GeV]
& $f$ [MeV] & $f^{\rm exp.}$\\
\tableline
$\pi$   & 0.22  & 0.3659  & 130 & 131\\
\tableline
$K$     & 0.45  & 0.3886  & 161.4 & 159.8$\pm$1.4\\
\tableline
$B$     & 5.2   & 0.5266  & 171.4 & 200$\pm$ 30~\protect\cite{Ber}
\end{tabular}
\label{t1}
\end{table}

\begin{table}
\caption{Results for form factors $F(0)$ and parameters
$\sigma_i$ defined in Eq.~\protect(\ref{Pole}).}
\begin{tabular}{c|ccc|ccc}
Model & $F_+(0)$ & $\sigma_1$ & $\sigma_2$
      & $F_T(0)$ & $\sigma_1$ & $\sigma_2$\\
\tableline
This work & 0.348 & 4.60E-2 & 5.00E-4 & $-0.324$&4.52E-2&4.66E-4 \\
\tableline
QM~\protect\cite{JW}&0.30 &6.07E-2&1.08E-3&$-0.30$&6.01E-2&1.09E-3\\
\tableline
QM~\protect\cite{Mel1}&0.36 &4.8E-2&6.3E-4&$-0.346$&4.9E-2&6.4E-4\\
\tableline
SR~\protect\cite{PB}& 0.341& 5.06E-2&5.22E-4&--&--&--\\
\tableline
SR~\protect\cite{AKS}& 0.35& 4.91E-2&4.50E-4&$-0.39$&4.91E-2&4.76E-4
\end{tabular}
\label{t2}
\end{table}

\begin{table}
\caption{Branching ratio(in units of
$|V_{ts}/V_{cb}|^2$) with[without] the pole contributions
for $B\to K\ell^+\ell^-$ for low, high, and total dilepton mass
region.}
\begin{tabular}{c|c|c|c}
Mode & $1\leq q^2\leq 8$  & $16.5\leq q^2\leq 22.9$ &
$ 4m^2_{\ell}\leq q^2\leq 22.9$ [GeV$^2$]\\
\tableline
$(e,\mu)$  & $2.59\times10^{-7}$ & $3.34\times10^{-8}$ & --\\
           & $[2.25\times10^{-7}]$ & $[3.70\times10^{-8}]$
           & $[4.96\times10^{-7}]$ \\
\tableline
$\tau$       & --   & $7.20\times10^{-8}$ & -- \\
             & --   & $[7.47\times10^{-8}]$ & $[1.27\times10^{-7}]$ \\
\end{tabular}
\label{t3}
\end{table}

\begin{table}
\caption{Non-resonant branching ratio(in units of
$10^{-7}\times|V_{ts}/V_{cb}|^2$) for $B\to K\ell^+\ell^-$
transition compared with other theoretical model predictions within
the SM as well as the experimental data taken from the Belle
Collaboration(by K. Abe et al.)~\protect\cite{Bfactory}.}
\begin{tabular}{cccccc}
Mode & This work &~\protect\cite{MN}&~\protect\cite{AKOS}
&~\protect\cite{ABHH}& Exp.~\protect\cite{Bfactory}\\
\tableline
$e$ & 4.96 & 4.4 & $3.2\pm 0.8$
& 5.7 & $<1.2\times 10^{-6}$\\
$\mu$ & 4.96 & 4.4 & $3.2\pm 0.8$
& 5.7 & $(0.99^{+0.39+0.13}_{-0.32-0.15})\times 10^{-6}$\\
$\tau$& 1.27 & 1.0 & $1.77\pm 0.40$ & 1.3&--
\end{tabular}
\label{t4}
\end{table}
\end{document}